\documentclass[preprint]{revtex4-1}

\usepackage{graphics}
\usepackage{graphicx}
\usepackage{pstricks}
\usepackage{amsbsy}
\usepackage{mathtools}
 
\renewcommand{\vec}{\mathbf}
\newcommand{\vecs}[1]{\boldsymbol{#1}}

\renewcommand{\tensor}{\mathbf}
\renewcommand{\div}{\boldsymbol{\nabla} \cdot}
\newcommand{\tensorStack}[2]{\stackrel{#1}{\mathbf{#2}}}
\renewcommand{\d}{\mathrm{d}}
\newcommand{\w}{\widetilde}
\newcommand{\hansenop}{\mathcal{H}}
\newcommand{\eg}{\textit{e.g.}}
\newcommand{\ie}{\textit{i.e.}}
\newcommand{\etal}{\textit{et al.}}
\newcommand{\seplib}{\texttt{seplib} }

\begin{document}
\title{Continuum Nanofluidics}

\author{J. S. Hansen}
\email{jschmidt@ruc.dk}
\affiliation{
  DNRF Centre ``Glass and Time'', IMFUFA, 
  Department of Science, , Systems and Models, 
  Roskilde University, Postbox 260, 
  DK-4000 Roskilde, Denmark
}

\author{Jeppe C. Dyre}
\affiliation{
  $\mbox{}$ DNRF Centre ``Glass and Time'', IMFUFA, 
  Department of Science, Systems and Models, 
  Roskilde University, Postbox 260, 
  DK-4000 Roskilde, Denmark
}

\author{Peter Daivis}
\affiliation{Applied Physics, School of Applied Sciences, RMIT University, GPO
  Box 2476, Melbourne, Victoria 3001, Australia
}

\author{Billy D. Todd}
\affiliation{
  Department of Mathematics,  Faculty of Science, Engineering and
  Technology, and Center for Molecular Simulation, 
  Swinburne University of Technology, PO Box 218, 
  Hawthorn, Victoria 3122, Australia 
}

\author{Henrik Bruus}
\affiliation{
  Department of Physics,
  Technical University of Denmark,
  DTU Physics Building 309,
  DK-2800 Kongens Lyngby,
  Denmark
}

\begin{abstract}
  This paper introduces the fundamental continuum theory governing
  momentum transport in isotropic nanofluidic flows. The theory is an
  extension to the classical Navier-Stokes equation, which includes
  coupling between translational and rotational degrees of freedom, as
  well as non-local response functions that incorporates spatial
  correlations. The continuum theory is compared with molecular dynamics
  simulation data for both relaxation processes and fluid flows
  showing excellent agreement on the nanometer length scale. We also
  present practical tools to estimate when the extended theory should
  be used. It is shown that in the wall-fluid region the fluid
  molecules align with the wall and in this region the isotropic
  model may fail and a full anisotropic description is necessary in
  order to describe this region.
\end{abstract}

\maketitle

\section{Introduction  \label{sec:intro}}
Nanoscale devices can now be fabricated with channels where the
smallest dimension is just a few nanometers \cite{eijkel_2005}, and
the development of nanofluidic theory
\cite{bocquet_2010,bruus_2008,eijkel_2005} is more relevant than
ever. Consider the following example. Perrson \etal~
\cite{persson_2007} used a series of rectangular nanochannels with
widths ranging from 14 to 300 nm to connect two micro-scale
chambers. By means of capillary filling, fluid from one chamber fills
up the channels and thus connects the two chambers. The filling rate
can be measured for different channel widths and for both milli-Q
water (filtrated de-ionized water) and an electrolyte solution of
sodium chloride. The rate did not follow the Washburn equation for
channel widths smaller than 100 nm.  The Washburn equation is based on
the classical continuum picture \cite{bruus_2008} using Poiseuille law
of fluid motion which includes the Newtonian (or macroscopic) shear
viscosity. For widths larger than 100 nm the Washburn equation
correctly predicts the filling rate. This is in accordance with the
common understanding that the discrete nature of the fluid at small
scales destroys the continuum picture
\cite{landau_1987,tritton_1988}. In fact, many researchers categorize
continuum physics as physics on the macroscopic scale, see for example
Ref. \onlinecite{lautrup_2005}.  Several questions immediately arise:
When exactly does the continuum picture fail? How is this breakdown
manifested? Does the length scale of the breakdown
depend on the specific problem? Can one improve the continuum
description such that it applies on small scales? 

\begin{figure}
  \scalebox{0.35}{
    \includegraphics{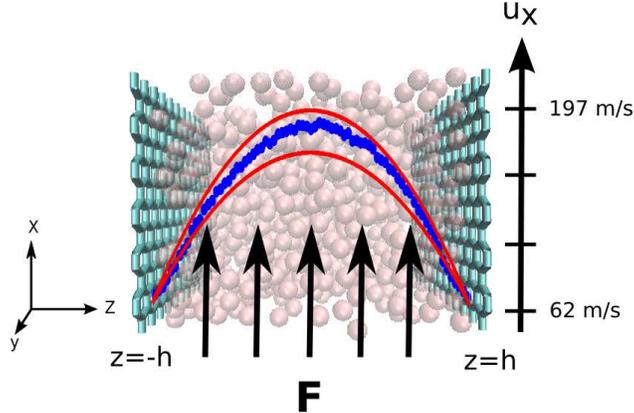}
  }
  \caption{ \label{fig:motivation} (Color online) Comparison between
    atomistic simulations (blue filled circles) and the continuum
    prediction (red lines) for a methane fluid undergoing a Poiseuille
    flow. The flow is generated by an external force field with
    magnitude $F=50$ TN, pointing along the $x$-direction. The
    Navier-Stokes equation predicts a velocity profile $u_x(z) = \rho
    F/(2\eta_0)(h^2-z^2) + u_w$, where $\rho=270$ kgm$^{-3}$ is the
    mass density, $\eta_0 = 9.3 \pm 0.6$ Pa$\cdot$s the Newtonian
    shear viscosity, and $u_w=62$ ms$^{-1}$ is the fluid slip velocity
    at the wall surface. The two lines represent the interval
    associated with the standard error in the viscosity
    \cite{rowley_1997}. The width of the slit pore is approximately 10
    molecular diameters or 3.3 nm.  }
\end{figure}

Demanding sufficient smoothness of the macroscopic quantities with
respect to time and position and using a simple statistical argument,
Lautrup \cite{lautrup_2005} estimates that the smallest volume
accessible to the continuum description must contain at least 10$^4$
molecules. This corresponds to a length scale of 8-80 nm, depending on
the density.  For steady flows the temporal fluctuations can be
averaged out and the accessible volume is much smaller.  This is
illustrated in Fig. \ref{fig:motivation}, where we have performed an
atomistic simulation (data given by blue filled circles) of a methane
fluid confined between two graphene sheets undergoing a Poiseuille
flow. The slit-pore has a width of approximately 3.3 nm and the flow
is driven by an external force field. In this paper the simulations
are carried out using the \seplib library \cite{seplib}. The classical
continuum prediction is plotted as two red lines illustrating the
maximum and minimum profiles allowed within statistical uncertainty on
the Newtonian shear viscosity \cite{rowley_1997}.  Only the fluid slip
velocity at the wall surface is used as a fitting parameter.
For this system the continuum theory gives a satisfactory description
of the fluid average velocity a length scale of a few
nanometers. Apparently, even on these small length scales the
molecular structure and degrees of freedom can be coarsened into
simple transport coefficients like the viscosity.  For water
undergoing a steady flow it has been shown by atomistic simulations
that the continuum description holds for channel widths of just 6-10
nm \cite{eijkel_2005,hansen_2011}. These results contrast earlier
assumptions about the validity of the continuum picture, and the
statement that continuum physics is physics on the macroscopic scale
\cite{lautrup_2005,landau_1987}. Interestingly, it was later argued by 
Thamdrup \etal~ \cite{thamdrup_2007} that the disagreement between the
experiment by Persson \etal~ \cite{persson_2007} and the Washburn
prediction is due to pinned micro-bubles resulting in an increase in 
hydraulic resistance.

At some point the classical continuum description will of course break
down. To mention two examples, Travis \etal~ \cite{travis_1997} showed
that for atomic fluidic systems the velocity profile features
modulations for confinements in the order of 5 atomic diameters.
Decheverry and Bocquet \cite{decheverry_2012} analyzed the effect of
thermal fluctuations on mass transport of fluid through a nanotube.
Interestingly, when the classical continuum theory fails, the dynamics
is frequently quantified by different transport coefficients compared
to those of the bulk system and effective transport coefficients are
introduced into the continuum constitutive relations
\cite{gubta_1998,karniadakis_2005}.

The main point of this paper is that the observation of a breakdown
need not be a failure of the continuum picture itself, but a result of
inadequate modeling wherein important dynamical processes are not
accounted for by classical theories. A very well understood example is
the effect of the Debye screening layer in electrolyte micro-flows
\cite{bruus_2008}.  Two other physical mechanisms that become
important on the nanoscale are often ignored in the literature, 
and this paper will treat these in detail:

(i) In classical hydrodynamics the fluid's local rotation is
determined uniquely by the fluid streaming velocity. One can quantify
the rotation from the local angular velocity field which is one half
the vorticity, that is, one half the curl of the streaming velocity
itself \cite{tritton_1988}. However, if the couple force, that is, the
force component producing pure rotation, is large, the rotation must
be treated as an independent variable. The extended description is
known as Cosserat (or micropolar) continuum mechanics
\cite{lukaszewicz_1999,eremeyev_2013}, first formulated by the
Cosserat brothers \cite{cosserat_1896,cosserat_1909} in the late 19th
century. Cosserat continuum theory is used in various areas such as
liquid crystal studies \cite{sonnet_2004} and blood flows
\cite{muthu_2008}, and was studied intensively in the 1950s to 1970s,
see
Refs. \onlinecite{grad_1952,dahler_1963,eringen_1969,degroot_1984,snider_1967,ailawadi_1971,evans_1978}. For
some reason it is not, however adopted by the nanofluidic
community. We show that Cosserat theory must also be used for fluid
flows in extremely small confinements where the molecular structure
becomes important.

(ii) Classical hydrodynamics is based on local constitutive models
relating fluxes to thermodynamic forces. For a shear flow the stress
at some point depends on the strain-rate at that particular point. If
the stress depends linearly on the strain rate, this leads to the
Newtonian law of viscosity \cite{tritton_1988}. A more general
constitutive relation is to let the stress be a function in the entire
strain rate history and spatial distribution, \ie, given by a spatial
and temporal convolution integral of a viscosity kernel and the strain
rate \cite{alley_1983}.  This is the approach of generalized linear
response theory \cite{boon_1991, evans_1990}.  The viscosity kernel
accounts for the characteristic length scale of the spatial
correlations \cite{furukawa_2009,puscasu_2010_3}; we show below that
this must be taken into account in order to arrive at the correct fluid
response on molecular length scales.

Our presentation is based on comparisons of continuum predictions with
atomistic molecular dynamics (MD) simulation data. These two
descriptions are fundamentally different in two ways. First, in MD the
system is characterized by discrete particles where the path of each
individual particle constituting the fluid is traced out through
classical mechanics \cite{allen_1989}, \ie, the particle interactions
must be known. The discretization of matter is, of course, in strong
contrast to the fundamental assumptions of continuum
mechanics. Secondly, the continuum description applies constitutive
relations to form mathematical closed problems. No such models are
enforced in the standard MD simulations.  Any discrepancy between MD
and the continuum description may therefore be a result of a breakdown
of the constitutive relation rather than a break down of the continuum
theory as such. Our basic conjecture is that MD acts as an idealized
numerical experiment, and if a given continuum theory agrees with the
MD data, the theory correctly accounts for the phenomena we study.

Let us specify, more accurately, what is meant by continuum
theory. Basically, one refers to deformable fluid volumes
characterized by quantities which are continuous at any point
$\vec{r}$ over the entire volume and at any time $t$
\cite{tritton_1988}. This means that these quantities are described
mathematically by field variables.  The basic continuum hypothesis is
that one can associate a given fluid sub-volume (or ``fluid
particle'') with the same characteristic quantities of the entire
deformable fluid volume, no matter how small the sub-volume
\cite{landau_1987, tritton_1988}. Lautrup \cite{lautrup_2005} suggests
a lower limit of the order of 10$^4$ molecules as stated above, but
time averaging allows an arbitrarily small fluid particle volume as
seen in Fig. \ref{fig:motivation}.  One field variable is the
streaming velocity, which is the mass-weighted average velocity of the
individual molecules in the fluid particle around a given point
\cite{bruus_2008}. The fluid's dynamics is governed by balance (or
conservation) equations. In general the balance equation for some
quantity per unit mass, $\phi=\phi(\vec{r},t)$, reads in the Eulerian
differential form \cite{degroot_1984}
\begin{equation}
\label{eq:generalbalance}
\frac{\partial (\rho \phi)}{\partial t} + \div \, (\rho
\vec{u}\phi) = \sigma_\phi - \div \, \tensor{J}_\phi \, ,
\end{equation}
where $\rho$ is the mass density, $\sigma_\phi$ a production term,
$\vec{u}$ the streaming velocity, and $\tensor{J}_\phi$ the flux of
$\phi$. Here $\phi$ can be a scalar or vector quantity. The right hand
side of Eq. (\ref{eq:generalbalance}) is the sum of the body force and
the surface force densities, that is, forces per unit volume. When
$\phi$ represent the velocity field, $\phi=\vec{u}$,
Eq. (\ref{eq:generalbalance}) is the momentum balance equation. The
body force density can be a gravitational-like force driving the flow
as in Fig. \ref{fig:motivation}, and the surface force density is the
pressure tensor, $\tensor{J}_\vec{u} = \tensor{P}$
\cite{evans_1990}. A special case is the mass balance equation for
which $\phi=1$. Since rotation is treated as an independent variable,
a balance equation on the form of Eq. (\ref{eq:generalbalance}) must
be formulated for rotation; this is done in the \emph{Supporting
Information}. (SI)  Importantly, in the extended Cosserat description the
pressure tensor $\tensor{P}$ need not be symmetric
\cite{grad_1952,degroot_1984,dahler_1963,eringen_1969} as in the
classical continuum theory. 

Comparison between the continuum description and MD simulation
data is carried out for molecular fluidic systems at equilibrium, as
well as for steady flows in a slit-pore. Here we investigate four
molecular fluids: a methane fluid, a generic di-atomic (dumbbell)
fluid, liquid butane, and liquid water. For methane 75 percent of the
mass is centered in the carbon nucleus and methane is here considered
as a simple spherical point-mass molecule, as it was done in
Fig. \ref{fig:motivation}. Water will, on the other hand, be treated
differently using the flexible SPC/Fw water model \cite{wu_2006} that
accounts for the molecular structure and hydrogen bonds and thus for
the structure of liquid water. The butane model is a coarse grained
model where the methyl and methylene groups are represented by a
united atomic unit, \ie, a spherical point-mass. Details about the
butane model can be found in Ref. \onlinecite{ryckaert_1978}, however,
here flexible bond are implemented with parameters from the
Generalized Amber Force Field \cite{wang_2004}. The simulations
are done using the \seplib library \cite{seplib}.

Nanofluidic flows are often associated with fluid slippage at the wall
boundary \cite{thomas_2008}. Just like the effect of the fluid-fluid
interactions on the flow is lumped into a single parameter, \eg,
viscosity, one effect from the fluid-solid interaction can be modelled
into a friction coefficient determining the boundary slip. The
slippage has a large effect on the flow rate in extreme confinement
and is usually quantified by the slip length $L_s$. For a
Hagen-Poiseuille flow in a tube with radius $R$ the relative flow
enhancement $\Delta E^{rel}$ due to the slip is given as
\cite{whitby_2008}
\begin{equation}
\Delta E^{rel} = 1 + 4L_s/R \ .
\end{equation}
$L_s$ is typically in the order of a few nanometers. Thus, for a given
non-zero slip length the flow enhancement increases hyperbolically as
the tube radius decreases. The slip is always present, but has
insignificant effect on the flow rate for tube radii above
microns. $L_s$ is normally independent of system size, that is, is not
an intrinsic nanofluidic phenomenon and is therefore not addressed in
this paper. Slippage is here modelled in an \emph{ad-hoc} fashion as
it was done in Fig. \ref{fig:motivation}. 

In SI we derive the Cosserat extended continuum
theory from the microscopic point of view using a microscopic
hydrodynamic operator. The derivation, which is based on the
fundamental definition of the macroscopic field variables in terms of
the corresponding molecular quantities, follows the idea of Irving and
Kirkwood \cite{irving_1950}, and Evans and Morriss \cite{evans_1990},
see also Ref. \onlinecite{todd_2007}. The
derivation leads to a molecular interpretation of the fluxes entering
Eq. (\ref{eq:generalbalance}). The final dynamical system of equations
is sometimes referred to as the extended Navier-Stokes (ENS) equations 
\begin{subequations}
\label{eq:sect1:ENS}
\begin{align}
  \rho \frac{\mathrm{D}\vec{u}}{\mathrm{D} t} &=
  \vecs{\sigma}_{\vec{J}} - \vecs{\nabla} p_{eq} + ( \eta_v + \eta_0/3
  - \eta_r) \vecs{\nabla} (\vecs{\nabla} \cdot \vec{u}) + (\eta_0
  +\eta_r) \nabla^2 \vec{u} + 2\eta_r \vecs{\nabla}\times\vecs{\Omega}
  \\ \rho I \frac{\mathrm{D} \vec{\Omega}}{D t} &=
  \vecs{\sigma}_{\vec{S}} + 2\eta_r(\vecs{\nabla} \times \vec{u} -
  2\vecs{\Omega}) + (\zeta_v + \zeta_0/3 - \zeta_r) \vecs{\nabla}
  (\vecs{\nabla} \cdot \vecs{\Omega})+ (\zeta_0 +\zeta_r) \nabla^2
  \vecs{\Omega} \ ,
\end{align}
\end{subequations}
where, $\mathrm{D}/\mathrm{D} t$ is the material operator,
$\vecs{\Omega}$ the spin angular velocity field, and $\vecs{\nabla} p$
is the pressure gradient. The transport coefficients $\eta_v$,
$\eta_0$, and $\eta_r$ are the bulk, shear, and rotational
viscosities, respectively, and $\zeta_v$, $\zeta_0$ and $\zeta_r$ the
corresponding spin viscosities. Finally, $\vecs{\sigma}_{\vec{J}}$ and
$\vecs{\sigma}_{\vec{S}}$ represent production terms of linear and
spin angular momentum, respectively. We refer the reader to SI for a
derivation and discussion of Eq. (\ref{eq:sect1:ENS}).

The theory is only strictly true for isotropic systems, and we study
such cases in Sections \ref{sec:coupling} and \ref{sect:nonlocal},
comparing theoretical predictions with MD simulation data. Flows in
extreme confinement are characterized by strong density
inhomogeneities and anisotropy. We study such flows in Section
\ref{sect:flows}, again comparing theory with MD data. Finally,
Section \ref{sect:summary} gives a brief summary. 

\section{Coupling: Multiscale relaxation phenomena in molecular fluids
  \label{sec:coupling}} 

The purpose of this section is to demonstrate the validity of the ENS
equations, Eqs. (\ref{eq:sect1:ENS}), by comparing the predictions of
different thermally induced relaxation phenomena with MD simulation
data, see also Refs. \onlinecite{hansen_2013, hansen_2013_2}. We start
with this problem instead of the situation with confining walls as the
latter is introduces density inhomogeneities and molecular alignment
at the wall-fluid interface. We return to this more complex situation
in Sect. \ref{sect:flows}.

Rather than investigating the quantities directly, one typically
studies the associated correlations \cite{kadanoff_2000}.  Here we
will use the approach based on Onsager's regression hypothesis
\cite{onsager_1931}, which states that thermal perturbations on
average decay according to the deterministic hydrodynamic equations of
motion. Specifically, we will compare mechanical spectra obtained from
MD simulations with predictions from theory.

\subsection{The stochastic ENS equations}
In equilibrium a fluctuating quantity $A$ can be written as
$A=A_{av}+\delta A$, where $A_{av}$ is the average part and $\delta A$
is the fluctuating part. In equilibrium the average streaming velocity
and spin angular velocity are both zero so $\vec{u} = \delta \vec{u}$
and $\vec{\Omega} = \delta \vecs{\Omega}$. The fluctuations are
modelled using the stochastic forcing approach
\cite{zarate_2006}. Here an uncorrelated zero mean stochastic force is
added to the constitutive relations, see SI. For example, for the
antisymmetric pressure the constitutive relation with stochastic
forcing reads $\tensorStack{ad}{P} = - \eta_r(\vecs{\nabla} \times
\delta \vec{u} - 2\delta \vecs{\Omega}) + \delta
\stackrel{ad}{\vec{P}}$, where $\delta \stackrel{ad}{\vec{P}}$ is the
fluctuating part of the flux.

To a first order approximation in the fluctuation we have on the
left-hand side of Eq. (\ref{eq:sect1:ENS})
\begin{equation}
(\rho_{av} + \delta \rho)\frac{\mathrm{D}\delta\vec{u}}{\mathrm{D} t}
\approx  \rho_{av} \frac{\partial \delta\vec{u}}{\partial t} \ \ 
\end{equation}
and
\begin{equation}
(\rho_{av} + \delta \rho) (I_{av} + \delta I) \frac{\mathrm{D}
  \delta \vecs{\Omega}}{\mathrm{D} t} \approx \rho_{av} I_{av}
\frac{\partial \delta  \vec{\Omega}}{\partial t} \ .
\end{equation}
In Fourier space the stochastic ENS equations read to first order in
the fluctuations for wave vector $\vec{k}$
 \begin{subequations}
  \begin{align}
    \rho_{av}\frac{\partial \w{\delta\vec{u}}}{\partial t} &= -
    i\vec{k}\w{p}_{eq} - (\eta_v +\eta_0/3 - 
    \eta_r)\vec{k}(\vec{k}\cdot\w{\delta \vec{u}}) -
     (\eta_0 +
    \eta_r)k^2\w{\delta \vec{u}} 
     + 
    2i\eta_r\vec{k}\times \w{\delta
      \vec{\Omega}} + i \vec{k} \cdot \w{\delta \tensor{P}}  
    \\
    \rho_{av} I_{av} \frac{\partial \w{\delta \vec{\Omega}}}{\partial t} =
    & 2\eta_r(i\vec{k}\times\w{\delta \vec{u}}- 2\w{\delta
      \vecs{\Omega}})  - 
    (\zeta_v + \zeta_0/3 - \zeta_r) \vec{k}(\vec{k}\cdot\w{\delta
      \vecs{\Omega}}) 
    -
     (\zeta_0 +\zeta_r)k^2\w{\delta \vecs{\Omega}} + 
     i\vec{k}\cdot  \w{\delta \tensor{Q}} + 2\w{\delta
      \tensorStack{ad}{P}} \ ,
  \end{align}
\end{subequations}
due to the properties of the divergence operator. See SI Eq. (2) for
the definition of the Fourier transform. It is here convenient to
introduce the following coefficients
\begin{equation}
  \eta_t = \eta_0 + \eta_r, \ \ \zeta_t = \zeta_0 + \zeta_r, \ \ \mathrm{and} \ \ \zeta_l = \zeta_v  +4\zeta_0/3 \ ,
\end{equation}
where subscripts $t$ indicates ``transverse'' and $l$ indicates
``longitudinal''. It has been shown \cite{hansen_2013,
  evans_1978} that $\zeta_t \approx \zeta_l$, and we write both
coefficients $\zeta$. We may then define the susceptibility
\begin{equation}
\chi(k^2) = 4\eta_r + \zeta k^2 \ ,
\end{equation}
where $k^2 = \vec{k}^2$. We will also drop the subscript $av$ from
here on.

We take $\vec{k}=(0, k_y, 0)$ and write out the $x$-component of the
velocity and $z$-component of the angular velocity
\begin{subequations}
  \label{eq:sect2:ens_u_k}
  \begin{align}
    \label{ens_du_k.1}
    \rho\frac{\partial \w{\delta u}_x}{\partial t} &=
    -\eta_tk^2\w{\delta u}_x
    + 2i\eta_rk_y\w{\delta\Omega}_z + ik_y\w{\delta P}_{yx} \\
    \label{ens_dw_k.1}
    \rho I \frac{\partial \w{\delta \Omega}_z}{\partial t} &=
    -\chi(k^2) \w{\delta\Omega}_z - 2i\eta_rk_y\w{\delta u}_x + ik_y\w{\delta Q}_{yz} +
   2 \w{\delta\stackrel{ad}{P}}_{z}  \ .
  \end{align}
\end{subequations}
These two components are both transverse components to the wave vector
and are coupled. We also investigate the longitudinal angular velocity
component $\w{\delta \Omega_y}$ which is given through
\begin{equation}
\rho{I}\frac{\partial \w{\delta \Omega}_y}{\partial t} =
-\chi(k^2) \w{\delta \Omega}_y + i k_y\w{\delta Q}_{yy} + 2
\w{\delta\stackrel{ad}{P}}_{y} .
\label{ens_dw_k.2}
\end{equation}
Note, that this longitudinal component is unaffected by the coupling
between the linear and spin angular momenta. 

We define the following three correlation functions
\begin{subequations}
  \label{eq:coupled-cf}
  \begin{align}
    C_{uu}^\perp(\vec{k},t) &= \langle \w{\delta u}_x(\vec{k}, t) \w{\delta u}_x(-\vec{k},
    0)\rangle/V \\
    C_{\Omega u}^\perp (\vec{k},t) &= \langle \w{\delta \Omega}_z
    (\vec{k}, t) \w{\delta  u}_x (-\vec{k}, 0) \rangle/V \\
    C_{\Omega \Omega}^{||} (\vec{k}, t) &= \langle
    \w{\delta \Omega}_y(-\vec{k},t) \w{\delta
      \Omega}_y(-\vec{k},0)\rangle/V  \ ,
  \end{align}
\end{subequations}
which we denote the transverse velocity auto-correlation function
(TVACF), the transverse cross-correlation function (TCCF), and
longitudinal angular velocity auto-correlation function (LAVACF),
respectively. By assumption the fluctuating fluxes are uncorrelated
with the velocity and angular velocity, \eg, $\langle \w{\delta
  P}_{yx} \w{\delta u_x}\rangle = 0$. Thus, multiplying
Eqs. (\ref{ens_du_k.1}) and Eq. (\ref{ens_dw_k.1}) with $\delta
u_x(-\vec{k},0)$ and ensemble averaging we arrive at the differential
equation system for the TVACF and the TCCF
\begin{subequations}
  \label{eq:sect2:cdff}
  \begin{align}
    \label{dtvacf.1}
    \rho\frac{\partial  C_{uu}^\perp}{\partial t} &=
    -\eta_tk^2 C_{uu}^\perp
    + 2i\eta_rk_y C_{\Omega u}^\perp\\
    \label{dtccf.1}
    \rho I \frac{\partial  C_{\Omega u}^\perp}{\partial t} &=
    -\chi(k^2)  C_{\Omega u}^\perp - 2i\eta_rk_y C_{uu}^\perp  \ .
 \end{align}
\end{subequations}
Similarly, multiplying Eq. (\ref{ens_dw_k.2}) with $\w{\delta
  \Omega}_y(\vec{-k,0})$ one has for the LAVACF 
\begin{equation}
\rho I \frac{\partial C_{\Omega \Omega}^{||}}{\partial t}
=   -\chi(k^2) C_{\Omega \Omega} \ 
\label{eq:delavacf}
\end{equation}
upon ensemble averaging. Now, Eqs. (\ref{eq:sect2:cdff}) and
(\ref{eq:delavacf}) can be solved yielding to second order in wave
vector 
\begin{subequations}
  \label{eq:corrfun}
  \begin{align}
    \label{eq:corrfun_2}
    C_{uu}^\perp(\vec{k},t) &=  \frac{k_BT}{4\rho} \left (
    Ik^2(-e^{-\omega_1 t} + e^{-\omega_2 t}) + 4 e^{-\omega_2
      t}\right) \\
    \label{eq:corrfun_3}
    C_{\Omega u}^\perp(\vec{k},t) &= - \frac{i 2 \eta_r k \left( e^{-\omega_1 t}
      -e^{-\omega_2 t} \right )}{4 \eta_r +(I(
      \eta_r - \eta_0) + \zeta)k^2} \\
    \label{eq:corrfun_1}
    C_{\Omega \Omega}^{||}(\vec{k},t) &= \frac{9k_BT}{4\rho I} e^{-\omega_0 t } ,
  \end{align}
\end{subequations} 
where the characteristic frequencies are
\begin{equation}
  \label{eq:eigen}
    \omega_0 = \frac{\chi(k^2)}{\rho I}, \ \ \ 
    \omega_1 = \frac{\chi(k^2) + I\eta_rk^2}{\rho I}, \ \ \
    \mbox{and}  \ \ \  
    \omega_2 = \frac{\eta_0 k^2}{\rho} \ .
\end{equation} 
The pre-factors in Eqs. (\ref{eq:corrfun_1}) and (\ref{eq:corrfun_2})
are calculated from the first order approximation in the
fluctuations $\vec{J} \approx \rho \delta \vec{u}$ as above and therefore
\begin{equation}
\w{\delta \vec{u}}(\vec{k},t) \approx \frac{m}{\rho}\sum_i \vec{c}_i
e^{-\vec{k}\cdot \vec{r}_i}
\label{eq:sigh1}
\end{equation}
from the definition of the linear momentum density, see SI 
Eq. (14) in SI, and for one component systems with molecular mass
$m$. Likewise, to a first order approximation in density and moment of
inertia $\rho \vec{S} \approx \rho I \vec{\Omega}$, and from SI
Eq. (29) 
\begin{equation}
  \w{\delta \vec{\Omega}}(\vec{k},t)
  \approx \frac{3m}{2\rho}\sum_i\vec{\Omega}_ie^{-\vec{k}\cdot
    \vec{r}_i} 
\label{eq:sigh2}
\end{equation}
as $I=2I_p/3$ \cite{sarman_1998}. Applying the equipartition
theorem one arrives at the pre-factors. Equations (\ref{eq:sigh1}) and
(\ref{eq:sigh2}) also provide a first order method to calculate the
correlation functions in the MD simulations; this method is used here.

It is informative to work in the frequency domain, \ie, to predict
the peak frequencies in the corresponding spectra. Applying the 
Fourier-Laplace transform defined by 
\begin{equation}
\mathcal{L}[f](\vec{k}, \omega) = \int_0^\infty f(\vec{k},t) \, 
e^{-i\omega t} \, \mathrm{d}t  
\end{equation}
we get 
\begin{subequations}
  \label{eq:sect2:corrfun}
  \begin{align}
    C_{uu}^\perp(\vec{k},\omega) &= -\frac{k_B T}{4\rho} \left(
       \frac{Ik^2}{\omega_1 + i\omega} - \frac{4 + Ik^2}{\omega_2 +
        i\omega} \right) 
     \label{eq:sect2:corrfun_2}\\
    C_{\Omega u}^\perp(\vec{k},\omega) &= 
 - \frac{i 2 \eta_r k}{\chi(k^2) + I(\eta_r - \eta_0)k^2}  
 \left( 
   \frac{1}{\omega_1 +i\omega} - \frac{1}{\omega_2 +i\omega}
 \right) ,
 \label{eq:sect2:corrfun_3} \\
    C_{\Omega \Omega}^{||}(\vec{k},\omega) &=    \frac{9k_BT}{4\rho I}
    \frac{1}{\omega_0 + i \omega} \ .
     \label{eq:sect2:corrfun_1} 
  \end{align}
\end{subequations} 

From Eq. (\ref{eq:sect2:corrfun_3}) we can make a very important
conclusion, namely,
\begin{equation}
    C_{\Omega u}^\perp(\vec{k},\omega) \rightarrow 0 \ \mbox{for} \
    \vec{k} \rightarrow \vec{0} \ .
\end{equation} 
This means that the coupling can be ignored on long length
scales. This is also expected as the classical
Navier-Stokes theory holds for macroscopic systems and no 
coupling effect is observed. The relaxation of spin is still
governed by the rotational viscosity, but this relaxation does not
affect the relaxation of linear momentum controlled by the usual
viscous dissipation processes. If we define $\omega_c$ as
\begin{equation}
\label{eq:wc}
w_c = \lim_{\vec{k} \rightarrow \vec{0}} \omega_0 =
\frac{4\eta_r}{\rho I},  
\end{equation}
the LAVACF and TVACF are, in the limit of zero wave vector,
\begin{equation}
 C_{\Omega \Omega}^{||}(\omega) = \frac{9k_B T}{4\rho I }
    \frac{1}{\omega_c + i \omega} 
\ \ \mbox{and} \  \
C_{uu}^\perp (\vec{k}, \omega) = \frac{k_B T}{\rho} 
\frac{1}{\omega_2 + i \omega}
\ \ \  (\vec{k} \rightarrow \vec{0}).
\label{eq:sect2:limitcase}
\end{equation}
Furthermore, for the fluids studied here the effect of the coupling on
the TVACF is not large, that is,
\begin{equation}
\left| \frac{Ik^2}{\omega_1 + i\omega}\right| < \left |\frac{4 +
    Ik^2}{\omega_2 + i\omega} \right|
\label{eq:unequiv}
\end{equation} 
even for wave vectors in the sub-molecular diameter range, and the
limit in Eq. (\ref{eq:sect2:limitcase}) need not to be taken as a
strict limit.  It is also worth noticing that the rotational viscosity
$\eta_r$ is a linear function of the moment of inertia $I$ for
sufficiently large $I$ \cite{moore_2008,hansen_2013_2}, so $w_c$ is
independent of $I$ here.

\subsection{Comparison with molecular dynamics}
We first compare the predictions from the continuum ENS theory with MD
simulation data for the simple di-atomic molecule (the dumbbell model)
in the super-critical fluid regime. The transport coefficients,
$\eta_0, \zeta$, and $\eta_r$ are listed in Table I in
SI.

\begin{figure}
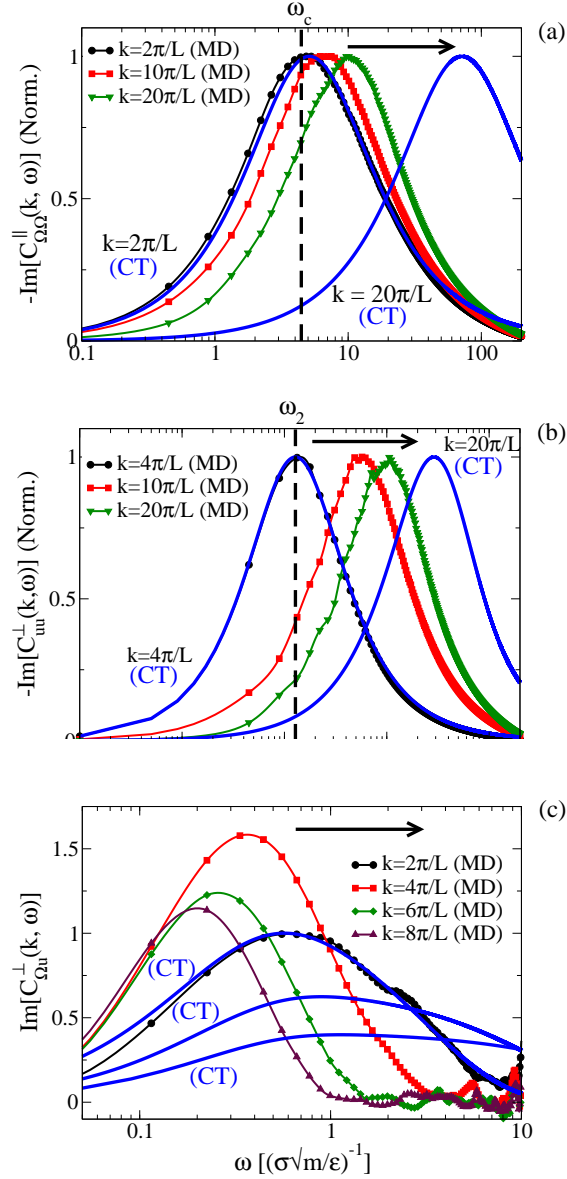

  \scalebox{0.275}{
    \includegraphics{DumbbellspectrumLAVACF.eps}
  }
  \\
  \scalebox{0.275}{
    \includegraphics{DumbbellspectrumTVACF.eps}
  }
  \\
  \scalebox{0.275}{
    \includegraphics{DumbbellTCCF.eps}
  }
  \caption{ \label{fig:spectraDumbbell} (Color online) Spectra for the
    dumbbell model. Blue lines are predictions from the continuum
    theory (CT) using the coefficients given in Table I, SI.
    The arrows indicate peak frequency behavior with increasing wave vector.
    (a) The imaginary part of the spectrum for the longitudinal
    angular velocity auto-correlation function (LAVACF). Normalization
    is carried out for clarity in the comparison. The dashed line
    indicates $\omega_c$ given by Eq. (\ref{eq:wc}) (b) As in (a), but
    for the transverse velocity auto-correlation function (TVACF).
    The dashed line indicates $\omega_2$ given by Eq. (\ref{eq:eigen})
    (c) As in (a), but for the transverse cross correlation function
    (TCCF). The predictions from the theory, Eq. (\ref{eq:corrfun_3}),
    for small wave vectors $k\leq 2\pi/L$ are shown.  
    Typical orders of magnitude for the MD units are $\sigma=1$ {\AA},
    $m = 10^{-26}$ kg and $\epsilon/k_B = 10^2$ K. $L=13.17 \sigma$.
  }
\end{figure}

Figure \ref{fig:spectraDumbbell} shows MD data (symbols connected with
lines) for imaginary parts of the spectra of the TVACF and LAVACF;
normalization is carried out for clarity. The prediction from the
continuum theory is plotted as full blue lines.
It is observed that for small wave vectors $k = 2\pi/L$ the continuum
prediction is in excellent agreement with the MD data, but it fails
for larger wave vectors $k = 20\pi/L$. We emphasize that no fitting is
performed, and all relevant parameters are taken from SI Table
I found from independent simulations and
methods. Using typical values for the MD units $\sigma, \epsilon$ and
$m$ the results show that the continuum theory predicts the mechanical
spectrum for wave lengths in order of 2-3 nm and above, and time
scales in order of of 1-10 ps and above.

For the TVACF, Fig. \ref{fig:spectraDumbbell} (b), the result can be
understood from the fluid stress relaxation at zero wave vector as
suggested by Bocquet and Charlaix \cite{bocquet_2010}. From the last
equation in Eq. (\ref{eq:eigen}) we can define a wave vector dependent
relaxation time $\tau_2 = 2\pi\rho/(\eta_0 k^2)$. This relaxation time
must be larger than the characteristic relaxation time $\tau_s$ at
zero wave vector for the predictions to hold $\tau_s < \tau_2$, \ie,
for the viscosity to be wave vector independent. This means that
\begin{equation}
\label{eq:ineqtvacf}
\frac{\eta_0 \tau_s k^2}{2\pi \rho} < 1 \ \ \mbox{or} \ \ k <
\sqrt{\frac{2\pi \rho}{\eta_0 \tau_s}} \ .
\end{equation} 
We will denote this the Bocquet-Charlaix criterion. Estimates for the
relaxation time $\tau_s$ is given through the shear pressure (or
equivalently stress) auto-correlator
\begin{equation}
G(t) = \frac{V}{k_BT}\langle \stackrel{os}{P}_{xy}(t)
\stackrel{os}{P}_{xy}(0) \rangle \ ,  
\end{equation} 
where $\stackrel{os}{P}_{xy}$ is the $(x,y)$ component of the
symmetric part of the pressure tensor. For the dumbbell model $G(t)$
is fully decayed at $\tau_s \approx 3 \sigma/\sqrt{m/\epsilon}$ which
gives $k < 1.2 \sigma^{-1}$. This is in perfect agreement with the
results depicted in Fig. \ref{fig:spectraDumbbell} (b). Alternatively,
the relaxation time can be given through the Maxwell relaxation time
$\tau_M = \eta_0/G^\infty$ or the viscous relaxation time
\cite{hartkamp_2013} $\tau_v = \Psi_{1,0}/2\eta_0$, where $G^\infty$
is the infinite shear modulus and $\Psi_{1,0}$ is the first normal
stress coefficient. For the diatomic model studied here $\tau_M
\approx \tau_v = 0.05 \sigma\sqrt{m/\epsilon}$ giving $k<9.4
\sigma^{-1}$ which is not what is observed. Therefore, the
characteristic decay time that should be used for the Bocquet-Charlaix
criterion is the time for the autocorrelation function $G(t)$ to fully
decay.

In the small wave vector regime the relaxation of spin angular
momentum is dominated by the coupling mechanism between linear and
angular momenta as the peak is located at $\omega_0 \approx \omega_c =
4\eta_r/(\rho I)$. The relaxation of linear momentum,
Fig. \ref{fig:spectraDumbbell} (b), is on the other hand due to usual
viscous mechanisms seen by the peak frequency $\omega_2 = \eta_0
k^2/\rho$. For large $k$ the continuum theory overestimates, by an
order of magnitude, the peak frequency for the LAVACF and TVACF due to
over-estimation of the effect of the spin diffusion.

Figure \ref{fig:spectraDumbbell} (c) depicts the TCCF for the dumbbell
model. Again, the theory performs surprisingly well for sufficiently
small wave vectors ($k \leq 2\pi/L$), but fails for larger. It is
worth noticing that the amplitude of the TCCF is a non-monotonic
function with respect to wave vector, having a maximum around
$k=4\pi/L$. This behavior is also captured by the ENS theory. To
illustrate that the amplitude is a decreasing function of wave vector
the TCCF for $k=\pi/L$ and $k=\pi/(2L)$ is plotted as predicted by the
theory. Recall, in the limit $\vec{k} \rightarrow \vec{0}$ the
coupling vanishes.

\begin{figure}
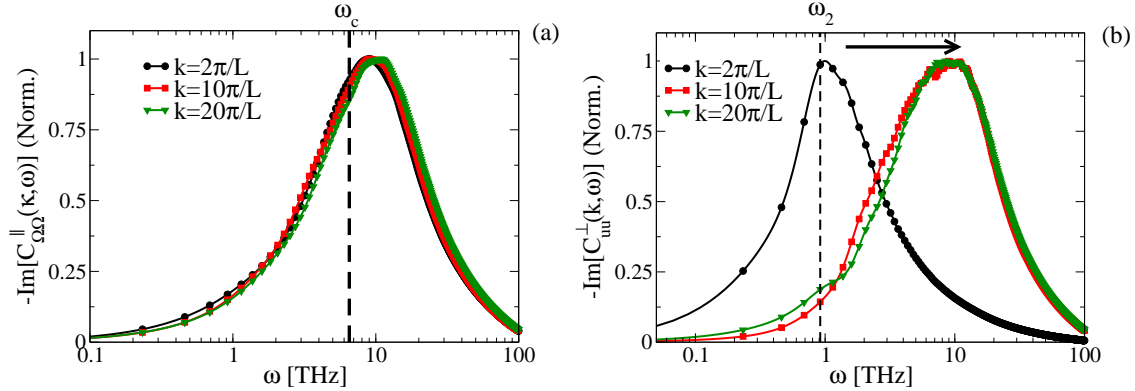

  \scalebox{0.275}{
    \includegraphics{ButaneSpectrumLAVACF.eps}
  }
  \scalebox{0.275}{
    \includegraphics{ButaneSpectrumTVACF.eps}
  }
  \caption{\label{fig:spectraButane} (Color online.) Liquid
    butane. Molecular dynamics results for (a) the LAVACF spectrum,
    and for (b) the TVACF spectrum.  The dashed lines indicate
    $\omega_c$, Eq. (\ref{eq:wc}), and $\omega_2$,
    Eq. (\ref{eq:eigen}). The arrow indicates peak frequency behavior
    with increasing wave vector. $L=32.95$ {\AA}.}
\end{figure}

Next we apply the theory to liquid butane. As discussed above the
butane model is not uni-axial or rigid, however, from the principal
moment of inertia we argued that the theory should be a good
approximation. The result is shown in Fig. \ref{fig:spectraButane}.
For the LAVACF one observes a peak frequency at around 9 THz and the
relaxation process is extremely fast. This fast mode is not precisely
captured by the theory with $w_c = 6.6$ THz. Interestingly, the peak
frequency is almost independent of wave vector for the range studied
here. This indicates that for these fast modes the diffusion of spin
is less important for the relaxation processes. For the slower
relaxation of linear momentum, we see that the peak frequency is
predicted very well by the theory, the MD result is $\omega$=1.0 THz
and the predicted one is around $\omega_2 = 0.92$ THz for the lowest wave
vector. Again, for larger wave vectors the prediction fails as expected.

\section{Non-local response \label{sect:nonlocal}}
The classical linear constitutive relations are local in the sense that
the flux only depends on the local and instantaneous thermodynamic
force. This is in general not the case, rather the response depends on
the entire force distribution in the system, as well as its
history. One can model this phenomenologically by introducing
frequency and wave vector dependent transport coefficients
\cite{evans_1990}. This renders the continuum description valid on
arbitrary small length and time scales. The generalized transport
coefficients are referred to as kernels. In the homogeneous isotropic
case, assuming space and time invariance, the linear non-local
constitutive relation for the symmetric part of the pressure tensor
reads \cite{evans_1990,puscasu_2010_1}
\begin{equation}
\label{eq:nonlocalconst}
\tensorStack{os}{P}(\vec{r}, t) = - \int_{-\infty}^t \int_{-\infty}^\infty
\eta(\vec{r}-\vec{r}', t-t') \dot{\vecs{\gamma}}(\vec{r}',t)
\mathrm{d}\vec{r}' \mathrm{d}t' \ .
\end{equation}
$\dot{\vecs{\gamma}}(\vec{r},t) = \tensorStack{os}{\nabla
  u}(\vec{r},t)$ is the symmetric part of the velocity gradient,
\ie, the strain rate. Fourier transforming with respect to space and
Fourier-Laplace transforming with respect to time yields
\begin{equation}
\w{\tensorStack{os}{P}}(\vec{k}, \omega) = - \w{\eta}(\vec{k}, \omega) 
\w{\dot{\vecs{\gamma}}}(\vec{k},\omega)
\end{equation}
from the convolution theorem. 

\begin{figure}
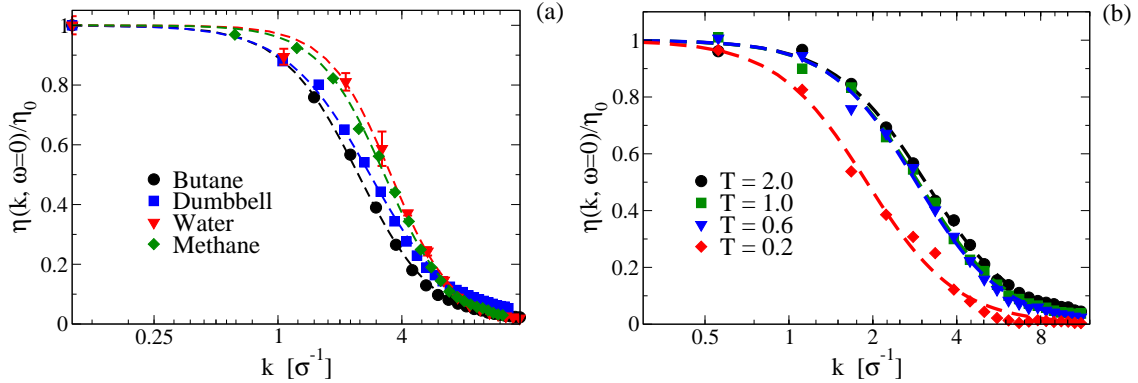

  \scalebox{0.275}{
    \includegraphics{kernel.eps}
  }
  \scalebox{0.275}{
    \includegraphics{kernel_asym.eps}
  }
  \caption{ \label{fig:kernel} (Color online) (a) Viscosity kernels
    for the dumbbell model, butane, water, and methane. For butane,
    water and methane $\sigma$ = 3.9233 {\AA}, 3.166 {\AA} and 3.80
    {\AA}, respectively. (b) Viscosity kernel for the asymmetric
    dumbbell model at different temperatures. The Newtonian viscosity
    values are $\eta_0 = 3.2 \sqrt{m \epsilon}/\sigma^2$ for $T=2.0
    \epsilon/k_B$ and $\eta_0 = 46 \sqrt{m \epsilon}/\sigma^2$ for
    $T=0.2 \epsilon/k_B$. For (a) and (b) the dashed lines are best
    fit to the empirical form $\w{\eta}(k) = \eta_0/(1 + \alpha
    k^\beta)$ \cite{hansen_2007_2} and is included to guide the eye.
  }
\end{figure}

The shear viscosity kernel $\w{\eta}(\vec{k}, \omega)$ can be found
from the TVACF as it is now shown. Here we focus on molecules with
small moment of inertia and small wave vector regime, \ie,  small
$Ik^2$. In this limit Eq. (\ref{eq:sect2:corrfun_2}) can be rearranged
giving 
\begin{equation}
\w{\eta}(\vec{k}, \omega) =   
\frac{ k_BT - i\omega \rho C_{uu}^\perp(\vec{k},
  \omega)}{k^2C_{uu}^\perp(\vec{k}, \omega)} \ \ \ \ (\mbox{small} \ Ik^2)
\ .
\label{eq:sect4:visckernel}
\end{equation}
In particular, we have at zero frequency
\begin{equation}
\label{eq:approxnonlocal}
\w{\eta}(\vec{k}, 0) = \frac{k_BT}{k^2C_{uu}^\perp(\vec{k},0)} \ \ \ \ 
\ \ \ \ \  \ \
(\mbox{small} \ Ik^2) .
\end{equation}
This approximation holds even for large values of $k^2$ as discussed
above, Eq. (\ref{eq:unequiv}). For molecules that can be regarded as
point masses, say methane, the moment of inertia is zero and
Eq. (\ref{eq:sect4:visckernel}) is exact.  In Fig. \ref{fig:kernel}
(a) the viscosity kernel at zero frequency is plotted for the methane,
dumbbell, butane, and water systems using Eq. (\ref{eq:approxnonlocal}).
One immediately notices that for $k \approx 1 \sigma^{-1}$ the wave
vector dependent viscosity approaches the zero wave vector limit. This
is in good agreement with the Bocquet-Charlaix criterion,
Eq. (\ref{eq:ineqtvacf}). Interestingly, this is independent of the
specific fluid studied here and the local constitutive relations can
be applied on length scales down to approximately $2\pi/k \approx$
2-2.5 nm.

Is this a general result that applies to all fluidic systems?  The
answer is no! In Fig. \ref{fig:kernel} (b) the zero-frequency
viscosity kernel is plotted for the asymmetric dumbbell model for
different temperatures.  The asymmetry arises due to the mass and
Lennard-Jones parameter differences between the two constituent
atoms. The asymmetric dumbbell model allows one to probe the dynamics
in the highly viscous regime without crystallization occurring
\cite{schroder_2009}. The result shows that for relatively high
temperatures the kernel has the same wave vector dependency, but
approaching the viscous regime (lower temperature) the kernel only
reaches the Newtonian viscosity at longer length scales. This
indicates that the dynamical processes behind the viscous response
take place on longer length scales in accordance with the cooperative
motion in super-cooled liquids \cite{cavagna_2009}. The non-local
viscous response has also been studied for highly viscous two
component Lennard-Jones system and polymer melts, see
Refs. \onlinecite{furukawa_2009, puscasu_2010_3}.

\begin{figure}
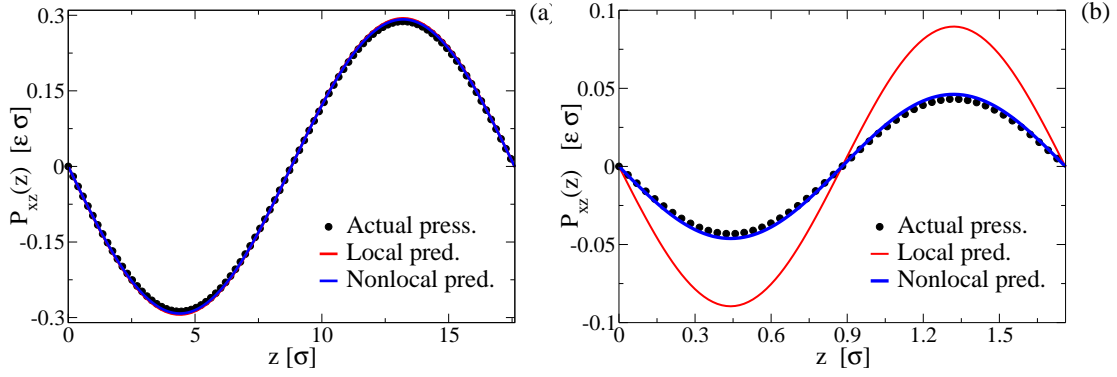

  \scalebox{0.275}{
    \includegraphics{localNonlocalStressLarge.eps}
    \includegraphics{localNonlocalStressSmall.eps}
  }
  \caption{ \label{fig:kernelStresses} (Color online) Pressure
    profiles for the Weeks-Chandler-Andersen point-mass system under
    periodic shearing force at state point $(\rho, T)=(0.685
    \sigma^{-3}, 0.765 \epsilon/k_B)$.  $\alpha = 4.81 \sigma^{-2}$ is
    found from best fit to data given in
    Ref. \cite{hansen_2007}. Input data for the theory is taken from
    Todd et al. \cite{todd_2008}.  (a) Small wave vector: $k=0.357
    \sigma^{-1}, \w{u}_x^k = 0.887 \sqrt{\epsilon/m}, F_0=0.15
    \epsilon/(\sigma m)$ (b) Large wave vector: $k=3.57 \sigma^{-1},
    \w{u}_x^k = 0.027 \sqrt{\epsilon/m}, F_0=0.225 \epsilon/(\sigma
    m)$ }
\end{figure}

The failure of the local constitutive relation, that is, of Newton's
law of viscosity, is very clearly illustrated by Todd \etal
~\cite{todd_2008} for a point-mass Weeks-Chandler-Andersen (WCA)
system \cite{weeks_1971}. In real space the non-local description
amounts to a convolution of the viscosity kernel and the strain rate
distribution, Eq. (\ref{eq:nonlocalconst}). In the homogeneous
situation where the fluid undergoes a steady shear in the
$x$-direction with varying amplitude in the $z$-direction we have one
non-zero shear component in the pressure tensor, namely, the $P_{xz}$
component. In this steady situation Eq. (\ref{eq:nonlocalconst})
reduces to
\begin{equation}
P_{xz}(z) = - \int_{-\infty}^\infty \eta(z-z')\dot{\gamma}(z') \,
\mathrm{d} z' \ ,
\label{eq:nonlocalz}
\end{equation}
where $\dot{\gamma}(z)=\partial u_x(z)/\partial z$. If the shear is
induced by an external force field $F_e(z) = F_0 \cos(k z)$, the fluid
flow is $u_x(z) = \w{u}_x^k\cos(k z)$, where $\tilde{u}_x^k$ is the
excited Fourier mode of the velocity field. We assume that this is the
only mode excited, \ie, the force amplitude must be sufficiently low
\cite{hansen_2007}. Also, this ensures a linear response as well as
constant temperature and density. The strain rate is then
\begin{equation}
\dot{\gamma}(z) = -k \w{u}_x^k \sin(k z) \ .
\label{eq:strainTodd}
\end{equation}
For simplicity we shall assume that the kernel is given by a Gaussian
function  
\begin{equation}
\eta(z)=\eta_0 \sqrt{\frac{\alpha}{\pi}} e^{-\alpha z^2} \ ,
\label{eq:kernelGaussian}
\end{equation}
such that $1/\sqrt{\alpha}$ gives a characteristic decay length. 
The kernel must fulfill \cite{todd_2008} (i) $\int_{-\infty}^\infty \eta(z)\,
\mathrm{d}z = \eta_0$, and (ii) $\eta(z)$ is an even function.
Substituting Eqs. (\ref{eq:strainTodd}) and (\ref{eq:kernelGaussian})
into Eq. (\ref{eq:nonlocalz}) we have upon integration
\begin{equation}
P_{xz}(z) = \eta_0 k \w{u}_x^k e^{-k^2/4\alpha} \sin(k z)\ .
\label{eq:stressNonlocal}
\end{equation}
If $\eta(z) = \eta_0\delta(z)$ the model is local, corresponding to
Newton's law of viscosity, that is, for the local model
\begin{equation}
P_{xz}^{L}(z) = -\eta_0 \dot{\gamma} = \eta_0 k \w{u}_x^k\sin(k z)
\ .
\label{eq:stressLocal}
\end{equation}
The system can be simulated using the Sinusoidal Transverse Force
(STF) method \cite{gosling_1973}, and it is possible to evaluate
$\w{u}_x^k$ for different external force fields and wave vectors.  The
two different predictions can be compared to the actual shear pressure
$P_{xz}^A$, which is found directly from the momentum balance equation
that for the steady flow reads
\begin{equation}
\frac{\partial}{\partial z}P_{xz}^A(z) = \rho F_e \ .
\end{equation}
Integrating we obtain 
\begin{equation}
P_{xz}^A(z) = \frac{\rho F_0}{k} \sin(k z) \ .
\end{equation}
The comparison is made in Fig. \ref{fig:kernelStresses}. Clearly, the
local prediction fails for the larger wave vector,
Fig. \ref{fig:kernelStresses} (b), whereas the non-local
prediction agrees with the actual shear pressure. From the non-local model
we conclude that spatial correlations result in a reduced shear pressure.

From Eqs. (\ref{eq:stressNonlocal}) and (\ref{eq:stressLocal}) we can
quantitatively evaluate the effect of spatial correlations on
the stress. Specifically, we have the relative difference given by
\begin{equation}
\label{eq:aargh1}
\Delta P_{xz}^{rel} = 1 - \frac{P_{xz}}{P_{xz}^{L}} = 1 - e^{-k^2/4\alpha} \ .
\end{equation}
For the WCA system studied here $\tau_s \approx 0.6
\sigma\sqrt{m/\epsilon}$ and the Bocquet-Chairlaix criterion gives $k
< 2.8 \sigma^{-1}$, this corresponds to an error in the stress below
32 \% according to Eq. (\ref{eq:aargh1}).

The Gaussian function does not perfectly fit to the kernel data.
Nevertheless, this simple functional form captures the non-local
response well, due to the smoothing of the convolution. Other more
complicated forms have been suggested, see
Refs. \onlinecite{hansen_2007_2, furukawa_2009,travis_1999}.

Todd and Hansen \cite{todd_2008_2} showed that the non-local response
is only relevant for flows where the strain-rate is non-linear with
respect to position. Couette and Poiseuille flows are then not affected
by non-locality. To illustrate this consider any functional form for
the kernel which fulfills the criteria given above: its integral gives
the zero wave vector viscosity and it is an even function with respect
to $z$. First, making the change of variables $u=z-z'$
Eq. (\ref{eq:nonlocalz}) reads
\begin{equation}
P_{xz}(z) = - \int_{-\infty}^\infty \eta(u)\dot{\gamma}(z-u) \,
\mathrm{d} u .
\end{equation}
Assuming a strain rate on the form $\dot{\gamma}(z) = \alpha z$, we
have 
\begin{equation}
P_{xz}(z) = - \alpha z \int_{-\infty}^\infty \eta(u) \, \mathrm{d} u 
+ \alpha \int_{-\infty}^\infty \eta(u) u \, \mathrm{d} u = - \alpha
\eta_ 0 z \ ,
\end{equation}
as the integrand in the second integral is an odd function. This
result is the same as the local prediction. In general, if a Taylor
expansion of the strain rate $\dot{\gamma} = a_0 + a_1 z + \ldots +
a_nz^n + \ldots$ exists, using the properties of odd and even
functions one can verify that the non-zero non-local effects of the
strain rate can be determined by the even moments of the kernel
\cite{todd_2008_2} 
\begin{equation}
M_n = \int_{-\infty}^\infty \eta(z)z^n \, \mathrm{d}z \ \ \ (n > 0 \
\mathrm{and \ even}) \ .
\end{equation}
In the case of Couette and Poiseuille flows the Taylor expansion
terminates at zero'th and first order, respectively, and there are no
non-local effects.

The spin and rotational viscosity kernels can be found by simply
rearranging Eq. (\ref{eq:sect2:corrfun_1}) giving the generalized
susceptibility
\begin{equation}
\w{\chi}(\vec{k}, \omega) = \frac{ k_BT
  - i\omega \rho I C_{\Omega \Omega}^{||}(\vec{k}, \omega)}{C_{\Omega
    \Omega}^{||}(\vec{k}, \omega)}  \ .
\end{equation}  
Our group recently \cite{hansen_2013, hansen_2013_2} conjectured that
the rotational viscosity $\eta_r$ governs the fast wave vector
independent relaxation processes as indicated in
Figs. \ref{fig:spectraDumbbell} (a) and \ref{fig:spectraButane}
(a). This transport coefficient is therefore only frequency
dependent. We then have
\begin{equation}
 \w{\chi}(\vec{k},\omega) = 4\w{\eta}_r(\omega) +\w{\zeta}(\vec{k}, \omega) k^2
\end{equation}
and therefore
\begin{equation}
\w{\eta}_r(\omega) = \frac{1}{4}\lim_{\vec{k}\rightarrow \vec{0}}  
\w{\chi}(\vec{k},\omega)
\ \ \mbox{and} \  \ 
\w{\zeta}(\vec{k}, \omega) = \frac{\w{\chi}(\vec{k},\omega) -
  4\w{\eta}_r(\omega)}{k^2}  .
\end{equation}
We called this the generalized extended Navier-Stokes (GENS)
theory. From MD simulations one can calculate the LAVACF (as shown
above) and from there find the kernels. For dense fluids $\w{\zeta}$
is characterized by a sharp peak around zero wave vector
\cite{hansen_2013} since the diffusive contribution to the relaxation
of the LAVACF is very small for $k > 2\pi/L$, see
Figs. \ref{fig:spectraDumbbell} (a) and \ref{fig:spectraButane}
(a). The spin viscosity kernel has the same properties as the shear
viscosity kernel and for this reason we do not expect any non-local effects
for flows where the gradient of the angular velocity is constant or
linear.
 
\section{Nanoflows \label{sect:flows}}

\subsection{The Poiseuille flow}
We first study a Poiseuille flow, the geometry is shown in
Fig. \ref{fig:motivation}.  In experiments this flow can be achieved
by application of a constant pressure gradient.  Generating a pressure
difference in simulations with, for example, a piston and using
molecular reservoirs can cause density variations in the direction of
the flow and other inlet/outlet effects. We therefore use a constant
force field acting on each point mass in the fluid to drive the flow.
The wall particles are arranged on a simple cubic lattice and are
allowed to vibrate around their initial lattice site using a simple
restoring spring force. The viscous heating generated in the fluid is
removed by thermostating the wall particles. This method resembles the
real physical experiment and is therefore often referred to as direct
non-equilibrium molecular dynamics. The interested reader is referred
to Ref. \cite{travis_1997} for further details. To get a satisfactory
signal-to-noise ratio in the MD simulations unrealistically large
external forces are typically applied to drive the system, and the
resulting flow rates are very large, typically, in the order of 10-100
m s$^{-1}$. Despite these large flow rates the Reynolds number is
usually less than unity due to the extremely small characteristic
length scales involved.  Finally, it is very important to ensure that
the simulations are carried out in the linear regime which is
discussed below.

\subsubsection{Continuum predictions}
In the linear regime of low Reynolds number and for the geometry shown
in Fig. \ref{fig:motivation} the ENS equations form a two-point
boundary value problem in the steady state
\begin{subequations}
  \begin{align}
    \rho F_e +\eta_t\frac{\d^2 u_x}{\d z^2} - 2\eta_r
    \frac{\d \Omega_y}{\d z} & = 0\\
    2\eta_r \left(\frac{\d u_x}{\d z} - 2 \Omega_y\right) +
    \zeta \frac{\d^2\Omega_y}{\d z^2} &= 0 \ ,
  \end{align}
\end{subequations}
where $-h\leq z \leq h$. Recall that $\eta_t = \eta_0 + \eta_r$ and
$\zeta = \zeta_0 + \zeta_r$. Introducing $z' = z/h, -1 \leq z' \leq 1$
and applying no-slip boundary conditions, $u_x(-1)=u_x(1)=0$
and $\Omega_y(-1)=\Omega_y(1)=0$, Eringen \cite{eringen_1969} solved
this yielding
\begin{subequations}
  \begin{align}
    u_x(z')&=u_c\left(
      1-z'^2 + \frac{2\eta_r\coth(Kh)}{\eta_tKh}
      \left(\frac{\cosh(Khz')}{\cosh(Kh)}-1\right)
    \right) 
    \label{eq:sect3:solu}
    \\
    \Omega_y(z') &= \frac{u_c}{h} \left(
       \frac{\sinh(Khz')}{\sinh(Kh)} - z'
    \right) \ ,
    \label{eq:sect3:solw}
    \end{align}
\end{subequations}
with the following definitions of $u_c$ and $K$
\begin{equation}
u_c = h^2\rho F_e/(2\eta_0) \ \ \mbox{and} \ \ 
K=\left(4\eta_r\eta_0/(\zeta\eta_t)\right)^{\frac{1}{2}}. 
\end{equation}
The application of the no-slip boundary condition is not justified. A
correct treatment applies the Neumann boundary condition for both the
velocity and angular velocity fields, however, this is not
straightforward in that the two are likely coupled and therefore
dependent on each other. While the boundary condition for the velocity
field has been studied in great detail, see \eg~ Refs.
\cite{navier_1823,bocquet_1994,hansen_2011_2}, very little is known
about the spin boundary condition. Just recently De Luca \etal~
\cite{deluca_2012} showed that the spin field does possess slippage
and Badur \etal~ \cite{badur_2015} used spin slip to account for flow
enhancement. As mentioned in the introduction, we will treat the
problem in an \emph{ad hoc} fashion and simply set the angular
velocity slip in accordance with the MD data.

If one ignores the coupling, $\eta_r=0$, the solution for
the streaming velocity, Eq. (\ref{eq:sect3:solu}), reduces to the
classical Poiseuille flow solution
\begin{equation}
\label{eq:pois}
  u_x(z')=u_c\left(1-z'^2\right) \, . 
\end{equation}
In this classical situation the angular velocity is found from the
vorticity, $\Omega_y = \frac{1}{2}\partial u_x/\partial z$, that is, 
\begin{equation}
\label{eq:sect3:solwClassical}
\Omega_y(z') =  -\frac{u_c}{h}z' = -\frac{h\rho F_e}{2\eta_o}z'
\end{equation}
in agreement with Eq. (\ref{eq:sect3:solw}) for $\eta_r \rightarrow
0$. As the classical treatment does not allow for specification of spin
boundary condition, Eqs. (\ref{eq:sect3:solw}) and
(\ref{eq:sect3:solwClassical}) differ by a magnitude of $h\rho
F_e/(2\eta_0)$ at the walls.

\begin{figure}
  \scalebox{0.275}{
    \includegraphics{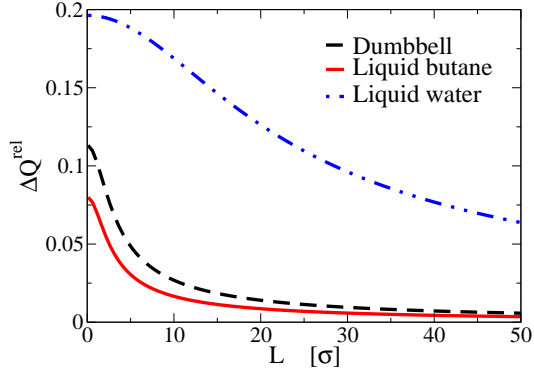}
  }
  \caption{ \label{fig:couplingEffect} (Color online) Relative
    volumetric flow rate reduction for the dumbbell fluid, liquid
    butane, and liquid water. $\sigma$ = 3.92 {\AA} and 3.17 {\AA} for
    butane and water, respectively.  }
\end{figure}

From Eq.  (\ref{eq:sect3:solu}) one can see that the maximum velocity,
located at $z'=0$, is lowered as a result of the coupling since from
the last term we have $1/\cosh(Kh)-1 < 0$ and thus $u_x(0) <
u_c$. Another way to quantify this effect is to evaluate the
volumetric flow rate $Q$ \cite{bruus_2008}
\begin{equation}
Q = \int_{-w}^{w} \int_{-h}^{h} u_x(z) \mathrm{d}z \mathrm{d}y =
2w\int_{-h}^{h} u_x(z) \mathrm{d}z \ ,
\end{equation}
where $w$ is the characteristic half length in the $y$-direction. This
gives the relative volumetric flow rate reduction
\begin{equation}
\Delta Q^{rel} = 1-\frac{Q}{Q_{\mathrm{class}}} = -\frac{3\eta_r(\tanh(Kh) -
  Kh)}{\eta_t\tanh(Kh)(Kh)^2} \ .
\label{eq:sect3:flowReduction}
\end{equation}
Equation (\ref{eq:sect3:flowReduction}) is plotted in
Fig. \ref{fig:couplingEffect} for the dumbbell fluid, liquid butane,
and water. The relevant coefficients can be found in SI Table I.  It
can be seen that for water flowing in a channel with a width of 9 nm
the flow rate is reduced by about 10 \% due to the coupling. As the
channel width increases the flow rate approaches that of the classical
predictions and the effect of the coupling can be ignored.

From Eq. (\ref{eq:sect3:flowReduction}) the relative flow rate
reduction increases as the product $Kh$ decreases.  From this
observation, one can define a characteristic fluid length scale $l_c$
\cite{hansen_2009_2} below which the effect of the coupling becomes
significant. To this end we write the parameter $K$ as
\begin{equation}
\label{eq:lcdef}
K = \frac{2}{l_c} \sqrt{\frac{\eta_0}{\eta_t}} \ \ \ \mbox{with}
\ \ \ l_c = \sqrt{\zeta/\eta_r}. 
\end{equation}
From SI Table I it is seen that $\eta_0 > \eta_r$ and $K
\approx 2/l_c$. Thus a significant flow-rate reduction occurs for
fluids with large critical length scale $l_c$. For water $l_c = 3.5$
nm and for butane $l_c = 0.5$ nm in agreement with the relative large
flow rate reduction observed for water.

\subsubsection{Comparison with molecular dynamics simulations}
First we compare MD data with continuum predictions for the dumbbell
system. Before the comparison, however, the linear Newtonian response
regime should be identified, at least, for the bulk fluid region. To this end one
can apply the synthetic SLLOD algorithm developed by Evans and Morris
\cite{evans_1984}. Basically the method imposes a constant strain rate
(linear velocity profile) on the system while ensuring a homogeneous
density and iso-kinetic temperature. To achieve this the equations of
motion are reformulated according to the Gaussian principle of least
constraint, see also Refs. \onlinecite{evans_1990,todd_2007}.
Performing a series of SLLOD simulations it is found that the
Newtonian regime occurs in the range $0<\dot{\gamma} < 0.05
(\sigma\sqrt{m/\epsilon})^{-1}$ for the dumbbell fluid . The upper
limit for the external force field can then be approximated by
rewriting Eq. (\ref{eq:sect3:solwClassical}) to $\dot{\gamma} \approx
2 \Omega_y = -2u_c/h z'$ giving $F_e < 2\eta_0\dot{\gamma}_m/\rho h$,
with $\dot{\gamma}_m=0.05 \sigma(\sqrt{m/\epsilon})^{-1}$. Note, in
the wall-fluid region the velocity may feature rapid changes and here
the linearity is not guaranteed.

Based on the SLLOD approach Delhommelle \cite{delhommelle_2002}
developed a synthetic method to calculate the rotational viscosity
$\eta_r$  as a function of spin angular velocity. See also Edberg
\etal~ \cite{edberg_1987_2}. To our
knowledge no synthetic or controlled method exists to study the spin
viscosity dependency of the gradient of the spin, or the rotational
viscosity dependency on strain rate. We will therefore here assume
that the linear regime is identical to the Newtonian regime, \ie, in
the regime where the viscosity is independent of the strain rate.

\begin{figure}
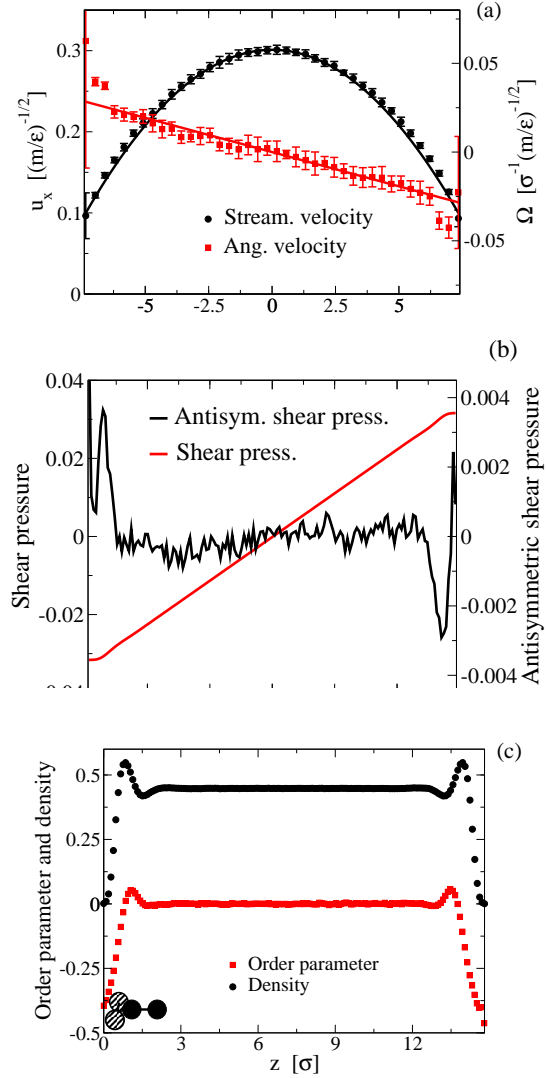

  \includegraphics[scale=0.25]{DumbbellProfVels.eps} \\
  \includegraphics[scale=0.275]{DumbbellProfPress.eps} \\
  \includegraphics[scale=0.25]{DumbbellProfOrderDens.eps} 
  \caption{\label{fig:prof-dumbbell} (Color online) (a) Velocity and
    angular velocity profiles for the dumbbell model undergoing a
    Poiseuille flow.  Symbols represent MD data, and lines the
    classical predictions where slippage is included.  (b) The
    corresponding shear pressures.  $P_{xz}$ is calculated by
    integration of the momentum balance equation, using the density
    profile given in (c). The antisymmetric pressure is calculated
    from the constitutive relation SI Eq. (40)
    using MD data as input.  (c) Density and order parameter
    profiles. The molecules in the lower left corner illustrate
    (exaggerated) the molecular ordering near the wall.  }
\end{figure}

The classical description predicts that the Poiseuille flow is local
flow according to Sec. \ref{sect:nonlocal}. Also, from
Fig. \ref{fig:couplingEffect} we expect the flow-rate reduction due to
the coupling to be very low for the dumbbell model. Thus, based on the
theory we can expect the classical description to be a good
approximation for this system. The time-averaged velocity and spin
angular velocity profiles are shown in Fig. \ref{fig:prof-dumbbell}
(a) for the dumbbell model where the pore width is approximately 14.8
atomic diameters. The profiles are sampled after the system has
reached the steady state. The temperature profile (not shown) is
constant and the temperature is $T = 4.0 \epsilon/k_B$ throughout the
channel.
The predictions from the classical Navier-Stokes theory,
Eqs. (\ref{eq:pois}) and (\ref{eq:sect3:solwClassical}), are also
plotted using the shear viscosity from SI Table I. Velocity slippage
at the wall-fluid interface is allowed, $u_x(z) = u_c(1-(z/h)^2) +
u_{w}$; $u_w$ is then the only fitting parameter in the
comparison. The agreement between MD data and classical continuum
predictions is excellent, except at the wall-fluid interface. This is
highlighted by the shear pressures plotted in
Fig. \ref{fig:prof-dumbbell} (b). According to the classical theory
the shear pressure $\stackrel{os}{P}_{xz}$ is linear, however, at the
wall-fluid interface this is not the case.  Also, the classical theory
assumes a zero anti-symmetric part of the shear stress
$\stackrel{ad}{P}_y$. This is clearly not fulfilled near the wall.

To understand the disagreement at the boundary, we analyze the fluid
ordering. It is well known that the wall induces a density variation
in the fluid \cite{toxvaerd_1981}. The density profile is shown in
Fig. \ref{fig:prof-dumbbell} (c) (black dots). It is seen that the
density varies in a region approximately one atomic diameter away from
the wall. The transport properties are functions of density, and one
should expect a variation in the viscosities here. Furthermore, one
can evaluate the molecular alignment ordering through the parameter
\cite{degennes_1993}
\begin{equation}
p = \frac{3}{2}\cos^2(\theta) - \frac{1}{2} \ ,
\end{equation}
where $\theta$ is the angle between the molecular bond and
$(x,y)$-plane. For perfect parallel alignment $p=-1/2$, that is, the
molecules closest to the wall are, on average, aligned with the
wall. For distances around one atomic diameter the molecules are
slightly normal to the wall as $p>0$. The extremes are illustrated
with the two molecules in the lower left corner in
Fig. \ref{fig:prof-dumbbell}(c). For zero order parameter, the
molecules have random orientations which is the case in the interior
of the channel. This means that the system possesses a degree of
anisotropy in the wall-fluid region. To fully account for the density
variation and ordering one should therefore describe the transport
properties through a position-dependent tensorial shear viscosity.

\begin{figure}
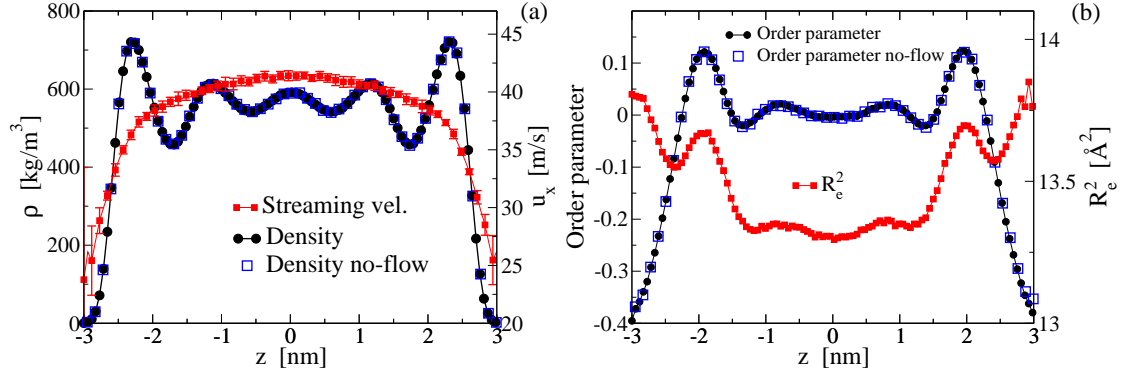

  \includegraphics[scale=0.275]{butaneSmall.eps}
  \includegraphics[scale=0.275]{butaneSmall-struct.eps}
  \caption{\label{fig:prof-butane} 
    (Color online.) (a) Density and velocity profiles
    for butane in a slit-pore of width 6 nm. The prediction from the
    theory breaks down and is not shown. (b) Corresponding order
    parameter and square end-to-end distance profiles.  }
\end{figure}

Figure \ref{fig:prof-butane} (a) shows the velocity and density
profiles for a butane flow where the pore width is just 6 nm. For such
extreme confinements the fluid layering stretches over the entire
pore. The order parameter profile, Fig. \ref{fig:prof-butane} (b),
shows that the molecular orientation is strongly anisotropic. Finally,
the mean square molecular end-to-end distance $R_e^2$ also varies,
showing that the butane molecule on average is elongated at the
fluid-region by around 2 per cent. Such a complex system is not
modeled appropriately by the classical or extended theories presented
here. This is not an indication of a breakdown of the continuum
picture, but an incomplete modeling. Worth noting is that the fluid
ordering and layering is constant over a large range of external
forces including zero force, see Fig. \ref{fig:prof-butane}, and is
thus not flow induced.

As pointed out by Bitsanis \etal~ \cite{bitsanis_1988} the velocity
profile features surprisingly small modulations considering the
density profile: one should expect the transport properties to vary
significantly across the channel having large effects on the flow
profile.  The authors suggested the local average density model (LADM)
wherein the transport properties at a point $z$ is a function of the
average density around that point. In the current geometry, where the
density is constant in the plane parallel to the wall, the local
average is
\begin{equation}
\overline{\rho}(z) = \frac{1}{\Delta}\int_{z-\frac{1}{2}\Delta}^{z+\frac{1}{2}\Delta}
\rho(z') \mathrm{d}z' \ ,\
\end{equation}
where $\Delta$ defines the region of averaging. The agreement between
the LADM and simulation data can be very good, especially if one
introduces a non-uniform weighting function
\cite{hoang_2012}. However, the LADM cannot predict the shear pressure
response in Fig. \ref{fig:kernelStresses} as the density is
constant. Also, the LADM model is not capable of predicting the strain
rate reversal observed by Travis \etal~ \cite{travis_1997, travis_2000}

To account for the observed velocity profile one can write the
position-dependent (in-homogeneous) non-local constitutive model as
\begin{equation}
  \stackrel{os}{P_{xz}}(z) = -\int_{-h}^h 
  \eta(z, z-z')\dot{\gamma}(z') \, \mathrm{d}z' \ .
  \label{eq:posnonlocal}
\end{equation} 
The position dependency likely comes from the varying density at the
wall-fluid region. The application of this relation is not
straightforward \cite{zhang_2004, cadusch_2008} as it is unclear how
the convolution should be performed at the wall where the support of
the kernel goes beyond the boundary and is unknown
\cite{zhang_2004,cadusch_2008}. Recently, Dalton \etal
\cite{dalton_2013} used a sinusoidal longitudinal force (SLF), also
introduced by Hoang and Galliero \cite{hoang_2012}, to control the
density variation in a periodic system. Due to the periodicity, the
boundary problem can be eliminated. The density profile can be
controlled to such an extent that it resembles that seen in confined
systems. The fluid can then be driven by an STF. The authors showed
that the non-local response is capable of predicting the strain rate
reversal observed by Travis \etal~ \cite{travis_1997, travis_2000}, as
well as the relative small modulation on the velocity profile. A
rigorous and general implementation of Eq. (\ref{eq:posnonlocal}) into
the balance equations for confined systems is still lacking.

\begin{figure}
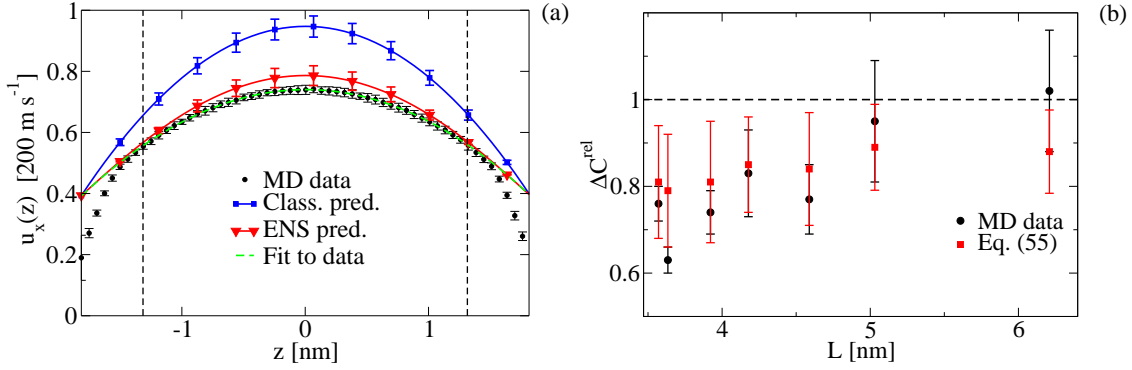

  \includegraphics[scale=0.275]{WaterVelProf.eps}
  \includegraphics[scale=0.275]{WaterCurveRel.eps}
  \caption{\label{fig:flow-water} 
    (Color online.)
    (a) Velocity profiles of water undergoing a Poiseuille
    flow.
    (b) Relative profile curvature difference at $z=0$.
    From Ref. \onlinecite{hansen_2011} with modifications. 
  }
\end{figure}

For the dumbbell and butane models the coupling between the linear and
spin angular velocities has little effect on the flow. From
Fig. \ref{fig:couplingEffect}, however, the effect is significant for
water flow in channel with widths below 5 nm. In
Fig. \ref{fig:flow-water} (a) MD data for the velocity profile for
water is plotted where the channel width is approximately 10 water
diameters.
Also, shown are the predictions from the classical NS and ENS
theories. Density and order parameter profiles (not plotted here) show
little density variation and molecular alignment, except within 3-4
{\AA} of the wall. The slip velocities are estimated by fitting a
second-order polynomial (dashed lines) to the velocity profile,
excluding the wall-fluid region where the fluid is slightly
anisotropic and inhomogeneous; this then amounts to the apparent slip
length \cite{lauga_2007} and is the only fitting parameter used in the
comparison. It is seen that the classical prediction fails, while the
ENS theory captures the flow profile Note, fit of the profile data to
the classical description will result in a wrong viscosity no matter
how many data points in the wall-fluid region are included. Any shift
of the profile will not change this either. An extra source of
dissipation must be present. Furthermore, note that a complete
description involves a position dependent non-local anisotropic
modelling of the wall-fluid region.

To remove any effect in the boundary region one can evaluate the
curvature in the channel midpoint, $z=0$. The predictions are simply
found from the second-order derivative of Eqs. (\ref{eq:sect3:solu})
and (\ref{eq:pois}). The relative difference is 
\begin{equation}
\Delta C^{rel} = 1 - \frac{\eta_r \coth(Kh)Kh}{(\eta_r + \eta_0) \cosh(Kh)} 
\ .
\end{equation}
$\Delta C^{rel}$ is plotted in Fig. \ref{fig:flow-water} (b) together with
the results from the MD simulations. Within statistical uncertainty
the ENS theory and MD simulation results agree. As the channel
width increases the relative curvature difference vanishes and the
classical description is re-captured. 

The particular model applied is parameterized with respect to the
liquid state and the wall is a Lennard-Jones cubic lattice, see
Ref. \onlinecite{hansen_2011}. The fluid structure near the wall and
its effect on the dynamics will be affected by the different
models, choice of model parameters, and wall details. However, it is not the
aim here to critically review the fluid structure at near the wall,
but to see the effect of the coupling.

\subsection{Inserting torque}
\begin{figure}
  \includegraphics[scale=0.275]{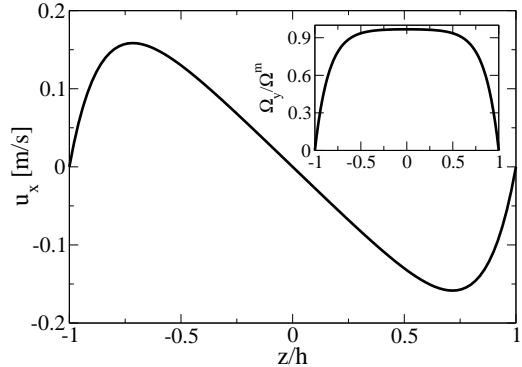}
  \caption{\label{fig:exttorque} 
    Flow of butane as a result applying an external torque with $h=$. The
    insert shows the corresponding angular velocity normalized with
    respect to the value in the limit of large $h$,
    Eq. (\ref{eq:largeh}), here denoted $\Omega_m$.  
  }
\end{figure}

Perhaps the most clear illustration of the translational-rotational
coupling is seen by introducing an external torque into the system
while having a zero production term for the linear momentum.  In
general, if the resulting torque density $\rho \Gamma_e$ is
sufficiently small then for the geometry in Fig. \ref{fig:motivation}
we have
\begin{subequations}
 \begin{align}
   \label{eq:torqueENSu}
   \eta_t\frac{\d^2 u_x}{\d z^2} - 2\eta_r
    \frac{\d \Omega_y}{\d z} & = 0\\
    \rho \Gamma_e + 2\eta_r \left(\frac{\d u_x}{\d z} - 
      2 \Omega_y\right) + \zeta \frac{\d^2\Omega_y}{\d z^2} &= 0 \ .
    \label{eq:torqueENSw}
 \end{align}
\end{subequations}
Integrating Eq. (\ref{eq:torqueENSu}) we get $\d u_x/\d z$ in terms of
$\Omega_z$ which is substituted into Eq. (\ref{eq:torqueENSw})
resulting in a second order inhomogeneous differential equation for
$\Omega_z$. From this and Eq. (\ref{eq:torqueENSu}) and by application
of Dirichlet no-slip boundary conditions one has
\begin{subequations} 
  \begin{align}
    u_x(z')      &=  \frac{4\eta_r}{\eta_tK} C_1 \sinh(Khz') - 
    \left(\frac{2\eta_r(\rho\Gamma\eta_t + 2\eta_rC_0)h}{\eta_t^2\zeta
      K^2} - \frac{C_0 h}{\eta_t}\right) z'  
    \label{eq:torqueSolu} 
    \\
    \Omega_y(z') &= 2C_1 \cosh(Kh z') -  \frac{\rho\Gamma \eta_t - 2\eta_r
      C_0}{\zeta\eta_t K^2} \ ,
  \end{align}
\end{subequations}  
where $C_0$ and $C_1$ are integration constants. One can show that 
$C_1$ goes rapidly to zero as $h$ increases. In this limit the 
spin angular velocity is 
\begin{equation}
\label{eq:largeh}
  \Omega_y(z') = -\frac{\rho\Gamma_e}{\zeta K^2} - 
  \frac{4\rho\Gamma_e\eta_r^2\eta_t K}{(\eta_t\zeta K^2)^2
    -4\eta_r\eta_t\zeta K^3} \,  ,
\end{equation}
and velocity profile is linear with a slope given by the
last term in Eq. (\ref{eq:torqueSolu}). Figure \ref{fig:exttorque}
depicts the two profiles for the butane liquid using $\Gamma_e = 413$
m$^2$s$^{-2}$.
From this one sees that the external torque produces a significant
local flow; the average flow is zero due to the system symmetry. 
  
In 2009 Bonthuis \etal~ \cite{bonthuis_2009} showed that the coupling
between the linear momentum and spin could be exploited in order to
pump water through carbon nanotubes by application of a rotating
field. The was theory based on the ENS equations at it was noted that
in order to obtain a non-zero mean flow asymmetric boundary conditions
must be employed which can be achieved by confining the fluid between
two walls with different hydrophobicity in the case of water pumping.
Recently De Luca \etal~ \cite{deluca_2012} performed extensive MD
simulations of the mechanism under experimentally feasible conditions,
indicating that the mechanism is functional. This could prove to be a
way to overcome the large hydraulic resistance characterizing
nanofluidic flows. Felderhof showed in 2011 that the coupling
can also be utilized to perform plane-wave pumping \cite{felderhof_2011}
and even propel microrobots \cite{felderhof_2011_4}.

\section{Summary \label{sect:summary}}
We have derived the relevant dynamical equations for
isotropic nanofluidic flows. The formulation is based on the
basic definition of a macroscopic field variable from the
corresponding microscopic or molecular variable, and it includes the
underlying molecular structure. Two intrinsic nanofluidic phenomena were
discussed, namely, (i) the coupling between the spin angular momentum and
(ii) the linear momentum and the non-local fluid response. The 
important points are the following.

(1) The effect of the coupling between the linear and spin angular
velocities can be estimated through the characteristic length scale
$l_c$, Eq. (\ref{eq:lcdef}). For large $l_c$ significant flow rate
reduction is observed, partly explaining the ``increased'' or
``effective'' viscosity reported in the literature. Other effects,
like anisotropy, will also play a role in the change in effective
transport properties.  For polar molecular systems like water $l_c
\approx 3-4$ nm, and the coupling must be considered on these length
scales. For the non-polar fluids studied here $l_c$ is below 1 nm and
the coupling effect is very small in most situations.

(ii) In general any fluid response can be described phenomenologically
through a transport kernel that incorporates the spatial and temporal
correlation effects. A method for calculating the shear viscosity
kernel was presented. This showed that for non-highly viscous fluids
the Newtonian limit is reached on length scales of a few
nanometers. This is in agreement with the Bocquet-Charlaix criterion
if the decay time for the stress autocorrelation function is applied
as the relaxation parameter. Importantly, non-local effects are
not present in simple flows where the strain-rate is linear with
respect to spatial coordinate, which is the case for Couette and
Poiseuille flows. For non-linear flows the non-local response
significantly affects the fluid stress for strain-rate variations 
on the atomic length scale. 
 
(iii) For highly confined fluids molecular alignment phenomena and
molecular deformation can occur along with the fluid layering, see
Fig. \ref{fig:prof-butane}. Simple classical continuum theory does not
include or account for such complex fluid structure. It would be
interesting to investigate this is more detail, for example, using 
the theory for liquid crystals \cite{degennes_1993,sarman_1993}. 

In conclusion, continuum theory is applicable even on the nanoscale 
if the relevant physical processes are modelled appropriately.

\section*{Acknowledgement}
The authors wish to acknowledge Lundbeckfonden for
supporting this work as part grant no. R49-A5634. The centre for
viscous liquid dynamics ``Glass and Time'' is sponsored by the Danish
National Research Foundation (DNRF).

\bibliographystyle{unsrt}

\section*{Supporting Information}

\section*{Balance equations and the Extended Navier-Stokes
  equation \label{sect:Nanofluidics}} 
A macroscopic field variable $A(\vec{r}, t)$ is be given by the
corresponding microscopic quantity $a_i(t)$ associated with molecule
$i$ through
\cite{hansen_1986}
\begin{equation}
\label{eq:sect1:micromacro}
A(\vec{r}, t) = \sum_i a_i(t)\delta (\vec{r}-\vec{r}_i(t)),
\end{equation}
where $\delta(\vec{r})$ is the Dirac delta function and $\vec{r}_i(t)$
is the molecular center-of-mass position.  In our treatment
$A(\vec{r},t)$ can be a scalar quantity (\eg~ the mass density) or a
vector quantity (\eg~ the momentum density). In Fourier space we have
for $A(\vec{r},t)$
\begin{equation} 
  \label{eq:Ftransform}
  \mathcal{F}[A(\vec{r},t)] \equiv 
  \widetilde{A}(\vec{k}, t) = 
  \iiint_{-\infty}^\infty  A(\vec{r},t) e^{-i\vec{k}\cdot
    \vec{r}} \, \mathrm{d} \vec{r} = 
  \sum_i a_i(t) e^{-i\vec{k}\cdot
    \vec{r}_i(t)} \, .  
\end{equation}
From now on we do not write the time and position dependencies of
the free variables (right hand side of the equations) explicitly unless
it provides important information. The rate of change in Fourier space
is given by
\begin{subequations}
  \begin{align}
    \frac{\partial}{\partial t} \w{A}(\vec{k},t) 
    &= \sum_i \left( 
      \frac{\d a_i}{\d t} + a_i (-i\vec{k}\cdot \vec{v}_i)
      \right) e^{-i\vec{k}\cdot\vec{r}_i} \\
    &= \sum_i \left( 
      \frac{\d a_i}{\d t} - i \vec{k} \cdot (\vec{v}_i a_i)
      \right) \left( 1 - i\vec{k}\cdot\vec{r}_i - \frac{1}{2}(\vec{k}\cdot
    \vec{r}_i)^2 \ldots \right) \ ,
    \label{eq:sect0:rateexpanded}
  \end{align}
\end{subequations}
where $\vec{v}_i$ is the center-of-mass velocity. The following
identity has been applied, where $a$ is a scalar or a vector 
\begin{equation}
\label{eq:dyadicindentity}
a(\vec{b}\cdot\vec{c}) = (\vec{b}\cdot\vec{c}) a = \vec{b} \cdot
(\vec{c} a) \ .
\end{equation} 
In the situation where $a$ is a vector the product
$\vec{c}\vec{a}$ is the dyadic between two vectors, $\vec{c}$ and
$\vec{a}$, and the resultant is the second rank tensor with components
$(ca)_ {ij}=c_ia_j$, $i,j=x,y,z$. The dyadic is not commutative, \ie,
$\vec{c}\vec{a} \neq \vec{a}\vec{c}$, unless $\vec{c}\vec{a}$ is
symmetric.  The dyadic is also called the outer product and sometimes
written as $\vec{c} \otimes \vec{a}$. If $\vec{c}$ and $\vec{a}$ are
parallel the dyadic is symmetric. To first order in wave vector
Eq. (\ref{eq:sect0:rateexpanded}) becomes
\begin{equation}
  \frac{\partial}{\partial t} \w{A}(\vec{k},t) = 
  \sum_i (1 - i\vec{k}\cdot\vec{r}_i) \frac{\d a_i}{\d t}
  - i\vec{k} \cdot \sum_i \vec{v}_ia_i  \ \ \ (\mbox{small}  \
  \vec{k}).
  \label{eq:sect:linear}
\end{equation}
The particle velocity can be decomposed into a thermal (or
''peculiar'') term $\vec{c}_i$ and advective term $\vec{u}(\vec{r}_i,
t)$ by
\begin{equation}
  \vec{v}_i(t) = \vec{c}_i(t) + \vec{u}(\vec{r}_i,t) \ .
\end{equation}
$\vec{u}(\vec{r}_i,t)$ is the mass averaged fluid velocity in a
region around $\vec{r}_i$. This region is equivalent to the fluid
particle volume encountered in the traditional continuum description
\cite{landau_1987} as was briefly discussed in
the introduction. We assume that the thermal and advective
velocities are uncorrelated and the thermal motion is, by definition,
conserved with $\sum_i m_i \vec{c}_i = \vec{0}$, where $m_i$ is the
mass. From this Eq. (\ref{eq:sect:linear}) becomes
\begin{equation}
  \label{eq:ratechangeA}
  \frac{\partial}{\partial t} \w{A}(\vec{k},t) = \hansenop [a_i] 
 \ \ \ (\mbox{small}  \ \vec{k}),
\end{equation}
where $\hansenop$ is a linear operator given by 
\begin{equation}
  \label{eq:mho}
\hansenop[a_i] = \sum_i (1-i\vec{k}\cdot \vec{r}_i)\frac{\d a_i}{\d
  t} - i\vec{k}\cdot \sum_i \vec{c}_ia_i - i\vec{k}\cdot \sum_i
\vec{u}(\vec{r}_i,t)a_i 
\ .
\end{equation}
$\hansenop$ is henceforth refered to as the ``microscopic hydrodynamic
operator'' (MH-operator) since it describes the microscopic
interpretation of the fluid's dynamics in the hydrodynamic regime of
small wave vectors. The first term describes the rate of change of the
microscopic quantity in question, the second and third terms are then
the thermal and advective contributions to the dynamics,
respectively. We proceed and apply Eqs. (\ref{eq:ratechangeA}) and
(\ref{eq:mho}) to express the balance equations of mass, linear
momentum and angular momentum.

\subsection{Mass balance}
The mass density field $\rho(\vec{r}, t)$ is defined directly from
Eq. (\ref{eq:sect1:micromacro}) as \cite{evans_1990}
\begin{equation}
\rho(\vec{r}, t) = \sum_i m_i\delta (\vec{r}-\vec{r}_i) \ .
\end{equation}
The MH-operator acting on $m_i$ gives
\begin{subequations}
  \begin{align}
    \hansenop [m_i](\vec{k},t) &=  \sum_i (1-i\vec{k}\cdot \vec{r}_i)\frac{\d m_i}{\d
  t} - i\vec{k}\cdot \sum_i m_i \vec{c}_i  - i\vec{k}\cdot \sum_i
m_i \vec{u}(\vec{r}_i,t)       
\\
    &= - i\vec{k}\cdot \sum_i m_i \vec{u}(\vec{r}_i,t)  
  \end{align}
\end{subequations}
since the mass of each particle is constant and $\sum_i m_i \vec{c}_i =
\vec{0}$. The dynamics in Fourier space is thus simply 
\begin{equation}
\frac{\partial}{\partial t} \widetilde{\rho}(\vec{k}, t) =  -
i\vec{k}\cdot \sum_i m_i \vec{u}_i   \ \ \ (\mbox{small}  \
  \vec{k}).
\label{eq:sect1:micromassbalance}
\end{equation} 
Relating to the general balance equation, Eq.(1) in the manuscript,
in the case of zero production term, one has $\phi = 1$, and no
surface forces 
\begin{equation}
\label{eq:sect1:massbalance}
\frac{\partial}{\partial t} \rho(\vec{r}, t) =  -\vecs{\nabla} \cdot
\rho \vec{u} \ . 
\end{equation} 
In Fourier space this yields 
\begin{equation}
\frac{\partial}{\partial t} \widetilde{\rho}(\vec{k}, t) = 
-i \vec{k} \cdot \widetilde{\rho \vec{u}}  
\end{equation}
using that $\mathcal{F}[{-\vecs{\nabla} \cdot (\rho \vec{u})}](\vec{k},t) =
-i \vec{k} \cdot \widetilde{\rho \vec{u}}$, which follows from partial
integration of Eq.(\ref{eq:Ftransform}). In the limit of small
wave vector we thus identify Eq. (\ref{eq:sect1:micromassbalance}) as
the microscopic interpretation of the continuous mass balance equation.

\subsection{Linear momentum}
In general, there is no microscopic definition of the streaming
velocity $\vec{u}$ in the form of Eq. (\ref{eq:sect1:micromacro})
\cite{evans_1990}. Rather, the linear fluid motion is given through
the momentum density $\vec{J}(\vec{r},t) = \rho(\vec{r},t)
\vec{u}(\vec{r},t)$. Note, Hansen and McDonald \cite{hansen_1986}
define a \emph{current} from the microscopic velocities and
Eq. (\ref{eq:sect1:micromacro}), which is then the correct mass
weighted averaged streaming velocity for single species fluids.  If
$ m_i \vec{v}_i$ is the linear momentum of particle $i$ we
have
\begin{equation}
  \label{eq:sect1:momentumcurrent}
    \vec{J}(\vec{r}, t) = \rho(\vec{r}, t) \vec{u}(\vec{r}, t) = 
    \sum_i m_i \vec{v}_i \delta(\vec{r} -\vec{r}_i) \ . 
\end{equation}
Applying the MH-operator we have
\begin{subequations}
  \begin{align}
    \hansenop [m_i\vec{v}_i](\vec{k},t) 
    &= 
    \sum_i (1-i\vec{k}\cdot \vec{r}_i)m_i\frac{\d \vec{v}_i}{\d t} 
    - i\vec{k}\cdot \sum_i m_i \vec{c}_i\vec{v}_i  - i\vec{k}\cdot \sum_i
    m_i \vec{u}(\vec{r}_i,t)\vec{v}_i   \\
    &=
    \sum_i (1-i\vec{k}\cdot\vec{r}_i) \vec{F}_i^{tot} -
    i\vec{k}\cdot \sum_i m_i \vec{c}_i \vec{c}_i - i\vec{k} \cdot
    \sum_i m_i\vec{u}(\vec{r}_i,t)\vec{u}(\vec{r}_i,t)  
    \label{eq:sect1:hydro_mom}
  \end{align}
\end{subequations}
as it can be shown \cite{todd_2007} that the cross terms $\sum_i
m_i \vec{u}(\vec{r}_i,t) \vec{c}_i = \sum_i m_i \vec{c}_i \vec{u}(\vec{r}_i,t) =
\tensor{0}$ since the thermal and advective velocities are
uncorrelated. 

$\vec{F}_i^{tot}$ is the total force acting on molecule $i$. This
force is decomposed into the force on $i$ due to interactions with all
other molecules denoted henceforth by $\vec{F}_i$ and external forces
$\vec{F}_i^{ext}$, \ie, $\vec{F}_i^{tot} = \vec{F}_i^{ext} +
\vec{F}_i$. The first term on the right hand side of
Eq. (\ref{eq:sect1:hydro_mom}) is
\begin{equation}
\sum_i (1-i\vec{k}\cdot{\vec{r}}_i)\vec{F}_i^{tot}  = 
\sum_i (1-i\vec{k}\cdot{\vec{r}}_i)\vec{F}_i^{ext} - i\vec{k}\cdot \sum_i
\vec{r}_i \vec{F}_i
\end{equation}
using Newton's third law $\sum_i \vec{F}_i = \vec{0}$ and the identity
Eq. (\ref{eq:dyadicindentity}).  Substituting
this into Eq. (\ref{eq:sect1:hydro_mom}) and rearranging one arrives
at
\begin{equation}
\frac{\partial}{\partial t} \widetilde{\rho\vec{u}}(\vec{k},t) 
+ i\vec{k}\cdot \sum_i m_i \vec{u}(\vec{r}_i,t) \vec{u}(\vec{r}_i,t)  = \widetilde{\vecs{\sigma}}_\vec{u}
- i\vec{k} \cdot \left(\sum_i m_i \vec{c}_i \vec{c}_i 
 + \sum_i  \vec{r}_i \vec{F}_i \right) \ \ \ \mbox{(small \ $\vec{k}$)},
\label{eq:sect1:momentumcurrent4}
\end{equation}
where 
\begin{equation}
\widetilde{\vecs{\sigma}}_\vec{u} = \sum_i (1 -
i\vec{k}\cdot\vec{r}_i) \vec{F}_i^{ext}.
\end{equation}
We recognize Eq. (\ref{eq:sect1:momentumcurrent4}) as the balance
equation for linear momentum in Fourier space. Importantly, the last
term on the right hand side gives \emph{one} possible microscopic
interpretation of the pressure tensor $\tensor{P}$ in the zero wave
vector limit \cite{todd_2007}.  Let $V$ be the system volume.  Then
according to Eq. (1) in the manuscript and
(\ref{eq:sect1:momentumcurrent4}) we have for the zero wave vector
pressure tensor
\begin{equation}
  \nonumber V\tensor{P}(t) \equiv \w{\tensor{P}}(\vec{k}=\vec{0}, t)
  = \sum_i m_i \vec{c}_i \vec{c}_i + \sum_i \vec{r}_i \vec{F}_i .
\end{equation}
This is the Irving-Kirkwood interpretation \cite{irving_1950} and will
be discussed in the following.

\begin{figure}[h]
  \begin{center}
    \scalebox{0.5}{
      \includegraphics{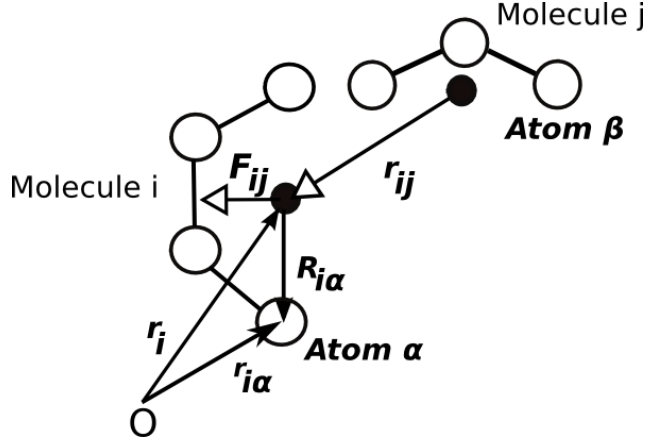}
    }
    \caption{\label{fig:sect1:antisymmetric} Schematic illustration
      of the interactions between two molecules $i$ and $j$. The atoms
      are represented by open circles and chemical bonds between atoms
      by straight lines. Filled circles are the center-of-masses. The
      force acting on $i$ due to $j$, $\vec{F}_{ij}$, needs not to be
      parallel with the vector $\vec{r}_{ij}$ connecting the two
      center-of-masses.  }
  \end{center}
\end{figure}

The configurational part of the pressure tensor is the term
$\sum_i  \vec{r}_i\vec{F}_i$ and can be written in a different form
assuming pairwise additive interactions only. Using Newton's third law
once again, $\vec{F}_{ij} = -\vec{F}_{ji}$, where $\vec{F}_{ij}$ is the
force on $i$ due to interactions with $j$, we get
\begin{equation}
\sum_i  \vec{r}_i\vec{F}_i 
= \sum_i  \vec{r}_i \sum_j  \vec{F}_{ij}
= \sum_i \sum_{j>i}   \vec{r}_{ij} \vec{F}_{ij} \, , 
\end{equation}
where $\vec{r}_{ij} = \vec{r}_i - \vec{r}_j$.  The Irving-Kirkwood
pressure tensor then reads
\begin{equation}
\label{eq:sect1:IK}
V\tensor{P}(t) = \sum_i m_i \vec{c}_i \vec{c}_i 
+ \sum_i \sum_{j>i} \vec{r}_{ij}\vec{F}_{ij} \, .
\end{equation}
At this point we emphasize the difference between the atomic and
molecular formalisms. If particles $i$ and $j$ represent point-masses
(including atoms in the molecules), rather than structured molecules,
the vectors $\vec{r}_{ij}$ and $\vec{F}_{ij}$ are parallel, the dyadic
$\vec{r}_{ij}\vec{F}_{ij}$ is symmetric. This implies that the
pressure tensor is symmetric. In the case of structured molecules,
however, the force on molecule $i$ due to $j$ is given by
\begin{equation}
\vec{F}_{ij}(t) = \sum_{\alpha \in i}\sum_{\beta \in j} \vec{F}_{i\alpha j\beta}
\, ,
\end{equation}
where indices $i\alpha$ and $j\beta$ represent atom $\alpha$ in molecule
$i$ and atom $\beta$ in molecule $j$, see
Fig. \ref{fig:sect1:antisymmetric}. The force $\vec{F}_{ij}$ needs not
be parallel to the vector $\vec{r}_{ij}$ connecting the molecules'
center-of-masses and the configurational part of the pressure tensor
is not generally symmetric.
There is no ambiguity between the two formalisms. They are trivially
equivalent for point-mass particles, and for structured molecules one
can be expressed in terms of the other and the mass dispersion tensor
\cite{allen_1984,todd_2007}. The two formalisms provide different
information and we use the molecular one here. We use the term 'atom'
loosely: it represents a point-mass spherical particle, and can be a
single atom or a united group of atoms like the methyl group.

In the molecular formalism the pressure tensor decomposes into a sum
of the symmetric $\tensorStack{s}{P}$ and anti-symmetric 
$\tensorStack{a}{P}$ parts such that $\tensor{P}=\tensorStack{s}{P} +
\tensorStack{a}{P}$ with
\begin{equation}
\label{eq:decompose}
\tensorStack{s}{P}(t)=\frac{1}{2}(\tensor{P} +
\tensor{P}^T) \ \ \mbox{and}
\ \ \tensorStack{a}{P}(t)=\frac{1}{2}(\tensor{P} - \tensor{P}^T) \, .
\end{equation}
Following Evans and Morriss \cite{evans_1990} we here use the stack
notation to indicate the symmetric and anti-symmetric parts. The
symmetric part can be further decomposed into a sum of the equilibrium
pressure $p_{eq}$, the viscous pressure $\Pi$, and the trace-less
symmetric tensor $\tensorStack{os}{P}$ \cite{evans_1990}
\begin{equation}
\tensorStack{s}{P}(t) = (p_{eq} + \Pi)\tensor{I} + \tensorStack{os}{P}
\ ,
\end{equation}
where $\tensor{I}$ is the second rank identity tensor and $p_{eq} +
\Pi=\mbox{Tr}(\tensor{P})/3$.  The anti-symmetric part of the
pressure tensor has zero diagonal components and three independent
off-diagonal components. It is therefore often written as a (pseudo)
vector dual $\tensorStack{ad}{P} = (\stackrel{a}{P}_{yz},
\stackrel{a}{P}_{zx}, \stackrel{a}{P}_{xy})$ of the tensor.  Here the
subscripts indicate the tensor components. From the definition of
$\tensorStack{a}{P}$, Eq. (\ref{eq:decompose}), one has for
$\tensorStack{ad}{P}$
\begin{eqnarray}
    2V\tensorStack{ad}{P}(t) &=& 
    \sum_i 
    (r_{y,i}F_{z,i} - r_{z,i}F_{y,i})\vec{i} + 
    (r_{z,i}F_{x,i} - r_{x,i}F_{z,i})\vec{j} + 
    (r_{x,i}F_{y,i} - r_{y,i}F_{x,i})\vec{k} \nonumber \\
    & =& \sum_i \vec{r}_i \times \vec{F}_i =  
    \sum_i \sum_{j>i} \vec{r}_{ij}  \times \vec{F}_{ij}  \ .
    \label{eq:vdualtorque}
\end{eqnarray}
Here $\vec{i}, \vec{j}$ and $\vec{k}$ are the unit vectors along the
$x$-, $y$- and $z$-axes, respectively.  From
Eq. (\ref{eq:vdualtorque}) it is clear that the anti-symmetric
pressure is associated with the torque on $i$ about $j$ and gives rise
to a change of orbital angular momentum, \ie, that of the
center-of-mass motion of the molecules. Importantly, when the pressure
tensor has a non-zero anti-symmetric part, the orbital angular
momentum is a non-conserved quantity \cite{degroot_1984}; we address
this in the following.  Conservation of orbital angular momentum is
sometimes used as an argument for a symmetric pressure, see for
example Aris \cite{aris_1962}, but note that later Aris also allows
for the possibility of asymmetry. Finally, notice that the kinetic
part of the pressure tensor $\sum_i m_i\vec{c}_i\vec{c}_i$ is a
symmetric dyadic and does not enter the anti-symmetric part.

\subsection{Orbital and spin angular momenta}
Consider now the case when no external forces are present. Following the
approach for the linear momentum we write the orbital angular momentum
density $\rho(\vec{r},t) \vec{L} (\vec{r},t)$ as 
\begin{equation}
  \rho(\vec{r},t) \vec{L} (\vec{r},t) = \sum_i \vec{L}_i
  \delta(\vec{r}-\vec{r}_i) = \sum_i (\vec{r}_i
\times \vec{p}_i) \delta(\vec{r}-\vec{r}_i) \, , 
\end{equation}
where $\vec{L}(\vec{r}, t)$ is the orbital angular momentum per unit
mass, and $\vec{L}_i = \vec{r}_i \times \vec{p}_i$ is the orbital
angular momentum of molecule $i$. Note, the angular momentum is
defined with respect to some specific choice of coordinate system. The
MH-operator acting on $\vec{L}_i$ gives
\begin{subequations}
  \label{eq:mho_L}
  \begin{align}
    \hansenop [\vec{L}_i](\vec{k},t) &= \sum_i (1-i\vec{k}\cdot \vec{r}_i)
    \frac{\d\vec{L}_i }{\d t} 
    - i\vec{k}\cdot\sum_i\vec{c}_i\vec{L}_i
    - i\vec{k}\cdot\sum_i\vec{u}(\vec{r}_i,t)\vec{L}_i 
    \\
    &=  \vec{N}  - i\vec{k} \cdot \sum_i \vec{r}_i\vec{N}_i 
    - i\vec{k} \cdot \sum_i \vec{c}_i \vec{L}_i - i\vec{k}\cdot
    \sum_i \vec{u}(\vec{r}_i,t)\vec{L}_i ,
    \end{align}  
\end{subequations}
where $\vec{N}_i = \vec{r}_i \times \vec{F}_i$ is the torque on
molecule $i$ and $\vec{N}$ is the sum of torques
\begin{equation}
  \vec{N} = \sum_i \vec{r}_i \times \vec{F}_i = 2 V
  \tensorStack{ad}{\tensor{P}} \ .
  \label{eq:totaltorque}
\end{equation}
The torque $\vec{N}$ does not sum to zero as the orbital angular
momentum is a non-conserved quantity \cite{degroot_1984}. Rather it is
the total angular momentum, \ie, the sum of orbital and spin (or
intrinsic) angular momenta, which is conserved.  For point-masses spin
angular momentum is not present (or meaningful) as there is no moment
of inertia. In this case the orbital angular momentum is conserved and
the torque $\vec{N}$ must be zero, that is,
$\tensorStack{a}{P}=\vec{0}$. The rate of change of orbital angular
momentum is then given by
\begin{equation}
\frac{\partial}{\partial t} \widetilde{\rho \vec{L}}(\vec{k}, t) + 
i\vec{k} \cdot \sum_i \vec{u}(\vec{r}_i,t)\vec{L}_i = 
 \vec{N}  - i\vec{k} \cdot \sum_i \left(\vec{c}_i \vec{L}_i+
\vec{r}_i\vec{N}_i \right)\ \ \ \mbox{(small \ $\vec{k}$)}. 
\end{equation}
This is the balance equation for the orbital angular momentum in
Fourier space.

The spin angular momentum density \cite{notQuantum} $\rho(\vec{r}, t)
\vec{S}(\vec{r}, t)$ is given by
\begin{eqnarray}
\label{eq:spindefinition}
\rho(\vec{r}, t) \vec{S}(\vec{r}, t) 
= \sum_i \vec{S}_i \delta (\vec{r}-\vec{r}_i)  = \sum_i \left(
\sum_{i\alpha} \vec{R}_{i\alpha} \times \vec{p}_{i\alpha}
\right)\delta (\vec{r}-\vec{r}_i) \, .
\end{eqnarray}
$\vec{S}(\vec{r}, t)$ is the spin angular momentum per unit mass,
$\vec{S}_i(t) = \sum_{i\alpha} \vec{R}_{i\alpha} \times
\vec{p}_{i\alpha}$ is the spin angular momentum of molecule $i$, and
$\vec{R}_{i\alpha}$ is the vector from the center-of-mass to atom
$\alpha$, see Fig. \ref{fig:sect1:antisymmetric}. Applying the
MH-operator and following the procedure in Eq. (\ref{eq:mho_L}) above,
the rate of change is to lowest order in wave vector
\begin{eqnarray} 
\frac{\partial}{\partial t} \widetilde{\rho \vec{S}}(\vec{k}, t) + 
i\vec{k} \cdot \sum_i \vec{u}_i \vec{S}_i =
\vec{M} - i\vec{k} \cdot \sum_i
\left(\vec{c}_i\vec{S}_i   + \vec{r}_i \vec{M}_{i}
\right)  \ \ \ (\mathrm{small} \ \vec{k}) ,
\label{eq:sect1:spinbalance}
\end{eqnarray}
where $\vec{M}_{i} = \sum_{i\alpha} \vec{R}_{i\alpha} \times
\vec{F}_{i\alpha}$ is the sum of torques on $i$ about the
center-of-mass and $\vec{M} = \sum_i \vec{M}_i$. Equation
(\ref{eq:sect1:spinbalance}) is the balance equation for the spin
angular momentum in Fourier space.

The last term in Eq. (\ref{eq:sect1:spinbalance}) defines the zero
wave vector Irving-Kirkwood couple tensor \cite{evans_1978}
$V\tensor{Q}(t) \equiv \tensor{Q}(\vec{k}=\vec{0}, t)$. The flux of
spin angular momentum associated with the couple tensor can be
important for flows in extreme confinements. After volume
averaging $\tensor{Q}$ can be written as
\begin{equation}
V\tensor{Q}(t) = \sum_i
\vec{c}_i \vec{S}_i + \sum_i \sum_{j>i}  \vec{r}_{ij}  \vec{M}_{ij} \ ,
\end{equation}
where
\begin{equation}
\vec{M}_{ij} = \sum_{i\alpha} \vec{R}_{i\alpha} \times \sum_{j\beta}
\vec{F}_{i\alpha j \beta} 
\end{equation}
is the torque, specifically the couple force, on molecule $i$ due to
interaction with all atoms $\beta$ in molecule $j$, see
Fig. \ref{fig:sect1:antisymmetric}. The couple tensor is not symmetric
and can be decomposed into symmetric and anti-symmetric parts just as
above for the pressure tensor, that is,
\begin{equation}
  \tensor{Q} = Q \tensor{I} + \tensorStack{os}{Q} + \tensorStack{a}{Q}
  \ .
\end{equation}
Also, one has a vector dual $\tensorStack{ad}{Q} =
(\stackrel{a}{Q}_{yz}, \stackrel{a}{Q}_{zx}, \stackrel{a}{Q}_{xy})$
for the anti-symmetric part of the couple tensor.

The total angular momentum is a conserved quantity.  This means that
in the limit of zero wave vector 
\begin{equation}
\frac{\partial}{\partial t} (\widetilde{\rho\vec{L}}
+ \widetilde{\rho\vec{S}}) = \vec{0}, \ 
\mbox{\ie}, \  \vec{M}  + \vec{N} = \vec{0} 
\label{eq:sect1:spinorbit}
\end{equation}
and so we have from Eqs. (\ref{eq:vdualtorque}) and
(\ref{eq:totaltorque})
\begin{equation}
\vec{M} = -\sum_i \vec{r}_i \times \vec{F}_i = -2V\tensorStack{ad}{P} \,  .
\end{equation}

Above we discussed the mass and momentum balance equations in the
framework of the microscopic picture, which is easiest done in Fourier
space. We will now return to the corresponding real space
formulations, but write the balance equations in a slightly different
form compared to Eq. (1) in the main manuscript. In real space
Eq. (\ref{eq:sect1:massbalance}) is the mass balance equation. Using
the product rule on the right hand side and re-arranging
\begin{equation}
\frac{\mathrm{D} \rho}{ \mathrm{D}t}  = 
- \rho (\vecs{\nabla} \cdot \vec{u}) \ ,
\label{eq:sect1:massbalanceReal}
\end{equation}
where, in general, the material derivative $\mathrm{D}/\mathrm{D t}$
is defined by
\begin{equation}
\frac{\mathrm{D} \phi}{\mathrm{D} t} =
\frac{\partial \phi}{\partial t} + \vec{u} \cdot ( \vecs{\nabla} \phi)
\ .
\end{equation}
Likewise, from Eqs. (\ref{eq:sect1:momentumcurrent4}) and
(\ref{eq:sect1:spinbalance}) we get the relevant momentum and spin
balance equations in real space after application of
Eq. (\ref{eq:sect1:massbalanceReal})
\begin{subequations}
  \label{eq:sect1:finalbalance}
  \begin{align}
    \rho \frac{\mathrm{D} \vec{u}}{\mathrm{D} t} &= \vecs{\sigma}_{\vec{u}} -
    \vecs{\nabla}\cdot((p_{eq} + \Pi) \vec{I} + \tensorStack{os}{P}) +
    \vecs{\nabla}\times \tensorStack{ad}{P} 
    \label{eq:sect1:finalbalancemom}\\
    \rho \frac{\mathrm{D} \vec{S}}{\mathrm{D} t} &=
    \vecs{\sigma}_{\vec{S}} -  2\tensorStack{ad}{P} - \vecs{\nabla}
    \cdot (Q \tensor{I} + \tensorStack{os}{Q}) + \vecs{\nabla} \times
    \tensorStack{ad}{Q} \ ,
    \label{eq:sect1:finalbalancespin}
  \end{align}
\end{subequations}
where $\vecs{\sigma}_{\vec{S}}$ represents the external production
term for the spin angular momentum.  In
Eqs. (\ref{eq:sect1:finalbalance}) we have applied the identity
$\vecs{\nabla} \cdot \tensorStack{a}{A} = -\vecs{\nabla} \times
\tensorStack{ad}{A}$ ($\tensor{A} = \tensor{P}$ or
$\tensor{Q}$). Interestingly, the term $-2 \tensorStack{ad}{P}$ in
Eq. (\ref{eq:sect1:finalbalancespin}) can be regarded as a production
term even in the absence of an external torque. Now,
Eqs. (\ref{eq:sect1:massbalanceReal}),
(\ref{eq:sect1:finalbalancemom}) and (\ref{eq:sect1:finalbalancespin})
give the relevant balance equations in the limit of small wave vectors
or, equivalently, large length scales.

Also, from Eqs. (\ref{eq:sect1:finalbalance}) one immediately sees that by
ignoring the anti-symmetric part of the pressure tensor we obtain the
classical momentum balance equation for the linear momentum. As
discussed above this applies to point-mass structure-less
fluids. Also, one observes that the dynamics of the spin angular
momentum is then determined by the couple forces alone and the spin is
a conserved quantity.

\subsection{The extended Navier-Stokes equations} 
The treatment above is general. We shall now focus on systems where
the molecules can be well approximated as uni-axial rigid body
molecules. Uni-axial molecules are defined as molecules where the
principal moment of inertia tensor has two nonzero components, 
denoted $I_p$, the third one being zero as the rotation around the main
molecular axis is not associated with any inertia.  This includes
di-atomic and linear molecules such as carbon dioxide, but other
molecules are also well approximated as uni-axial and rigid, for
example, butane as we will see below. The treatment in this section is
not as detailed as above, and we refer the reader to
Refs. \cite{degroot_1984, sarman_1998,travis_1997_3, delhommelle_2002_2}. 

It is convenient to describe the spin dynamics in the principal frame
of reference where the moment of inertia tensor is constant. From the
Euler equation of motion for rigid bodies \cite{goldstein_2002} one
can show that in this frame and for uniaxial molecules the left hand
side of Eq. (\ref{eq:sect1:finalbalancespin}) becomes
\begin{equation}
\rho \frac{\mathrm{D}\vec{S}}{\mathrm{D} t} = \rho I
\frac{\mathrm{D}\vecs{\Omega}}{\mathrm{D} t}   \ ,
\label{eq:sect1:angularvel}
\end{equation}
where $I = 2I_p/3$ \cite{sarman_1998, travis_1997_3} and
$\vec{\Omega}$ is the angular velocity. Note, this does
not hold in general where nonlinear coupling between the angular
velocity components are present. The general situation, however, 
can be included into the theory.

For isotropic systems the following linear local 
constitutive relations are applied, see
Refs. \onlinecite{evans_1978,degroot_1984}, 
\begin{subequations}
\label{eq:sect1:constrelations}
\begin{align}
  \Pi &= -\eta_v (\vecs{\nabla} \cdot \vec{u}) \\
  \tensorStack{os}{P} &= -2 \eta_0  (\stackrel{os}{\vecs{\nabla}
    \vec{u}}) = - \eta_0 \left( \left(\vecs{\nabla}\vec{u} +
      \vec{u}\vecs{\nabla}\right) -
    \frac{2}{3}\mathrm{Tr}(\vecs{\nabla}\vec{u}) \tensor{I}
  \right) 
  \\
  \tensorStack{ad}{P} &= -\eta_r (\vecs{\nabla} \times \vec{u} - 2
  \vecs{\Omega}) \label{eq:antistressConst}
  \\
  Q &= -\zeta_v (\vecs{\nabla} \cdot \vecs{\Omega}) \\
  \tensorStack{os}{Q} &= -2 \zeta_0  (\stackrel{os}{\vecs{\nabla}
    \vec{\Omega}}) 
= - \zeta_0 \left( \left(\vecs{\nabla}\vecs{\Omega} +
      \vecs{\Omega}\vecs{\nabla}\right) -
    \frac{2}{3}\mathrm{Tr}(\vecs{\nabla}\vecs{\Omega}) \tensor{I}
  \right)  
  \\
  \tensorStack{ad}{Q} &= -\zeta_r (\vecs{\nabla} \times \vec{\Omega} )
  \ .
\end{align}
\end{subequations}
Here $\eta_0, \eta_v$ and $\eta_r$ are the shear, bulk and rotational
viscosities, respectively, and $\zeta_0, \zeta_v$ and $\zeta_r$ are
the corresponding spin viscosities. Usually, a thermodynamic force is
defined through the gradients of the field variables, however, in
Eq. (\ref{eq:antistressConst}) we recognize the thermodynamic force
$\vecs{\nabla}\times\vec{u} - 2 \vecs{\Omega}$ as twice the difference
between the classical prediction of the angular velocity,
$\vecs{\Omega} = \frac{1}{2}\vecs{\nabla}\times \vec{u}$, and the
actual one. This force is denoted the sprain rate \cite{evans_1978}.
Substituting Eqs. (\ref{eq:sect1:constrelations}) into
Eqs. (\ref{eq:sect1:finalbalance}) and using
Eq. (\ref{eq:sect1:angularvel}) we arrive at the extended
Navier-Stokes (ENS) equations \cite{evans_1978}
\begin{subequations}
\begin{align}
  \rho \frac{\mathrm{D}\vec{u}}{\mathrm{D} t} &=
  \vecs{\sigma}_{\vec{J}} - \vecs{\nabla} p_{eq} + ( \eta_v + \eta_0/3
  - \eta_r) \vecs{\nabla} (\vecs{\nabla} \cdot \vec{u}) + (\eta_0
  +\eta_r) \nabla^2 \vec{u} + 2\eta_r \vecs{\nabla}\times\vecs{\Omega}
  \\ \rho I \frac{\mathrm{D} \vec{\Omega}}{D t} &=
  \vecs{\sigma}_{\vec{S}} + 2\eta_r(\vecs{\nabla} \times \vec{u} -
  2\vecs{\Omega}) + (\zeta_v + \zeta_0/3 - \zeta_r) \vecs{\nabla}
  (\vecs{\nabla} \cdot \vecs{\Omega})+ (\zeta_0 +\zeta_r) \nabla^2
  \vecs{\Omega} 
\end{align}
\label{eq:sect1:ENS-supplementary}
\end{subequations}
using the divergence rule $\vecs{\nabla} \cdot
(\tensorStack{s}{\vecs{\nabla}\vec{u}}) = \frac{1}{2} \nabla^2
\vec{u} + \frac{1}{2} \vecs{\nabla}
(\vecs{\nabla}\cdot\vec{u})$. 

We now present values for the transport coefficients entering
Eqs. (\ref{eq:sect1:constrelations}) for the molecular fluids studied
here. Methane is modelled as a point-mass molecule where the coupling
is irrelevant, \ie, we let $\eta_r = 0$. Butane and water are not
uni-axial. For the butane model used the principal moment of inertia
components are $I_x=2.2 \times 10^{-20 }$ m$^2$, $I_y=1.8 \times
10^{-20}$ m$^2$ and $I_z=0.26 \times 10^{-20 }$ m$^2$, that is, the
molecule has one major axis and two almost identical minor axes and we
can expect the theory to hold reasonably well. In
Ref. \onlinecite{hansen_2010} the water molecule was considered as a
dipolar rotator with an effective moment of inertia of $I=8.4 \times
10^{-22}$ m$^2$. This approach will be adopted here. In Table
\ref{table:coeff} the relevant state points and transport coefficients
are listed. In the treatment below the dynamics is decomposed into
transverse shear and longitudinal bulk modes. For linear momentum the
focus is on the shear mode, and the bulk viscosity is therefore not
listed in the table. The coefficients are evaluated from independent
equilibrium MD simulations as prescribed in
Refs. \onlinecite{allen_1989,evans_1978,moore_2008,hansen_2010}.  The
values for the dumbbell model are given in reduced MD units of length
scale $\sigma$, energy scale $\epsilon$, and mass $m$, see
Ref.\onlinecite{allen_1989}.
\begin{table}[h]
\begin{tabular}{ccccccccccccc}
\hline \hline
Molecule && $\rho$ [kg m$^{-3}$] && T [K] && 
$\eta_0$ [mPa$\cdot$s] && $\eta_r$ [mPa$\cdot$s]  && 
$\zeta_0 + \zeta_r$ [kg m s$^{-1}$] && I [m$^2$]\\
\hline
Methane   && 460    && 164.4 K && 0.27$^*$ && - && - && - \\
Dumbbell$^{**}$  && 0.4477 && 4.0     && 0.60     && 0.083  && 0.22 && 1/6 \\
Butane    && 582.3  && 288 K   && 0.14     &&  0.013
&& 4.0 $\times$ 10$^{-24}$ && 1.3 $\times$10$^{-20}$ \\
Water     && 996.3 && 298.7   && 0.7  && 0.17 && 
2.1 $\times$ 10$^{-21}$ && 8.4 $\times$ 10$^{-22}$  \\
 \hline \hline
\end{tabular}
\caption{\label{table:coeff} 
  $^*$ From Ref. \onlinecite{rowley_1997}. $^{**}$ Quantities given in
  reduce MD units. \\ 
  State points, transport coefficients and moments of inertia for the
  systems studied. 
}
\end{table}

\bibliographystyle{unsrt}

\begin{thebibliography}{10}
\bibitem{notQuantum}
This is not to be confused with particle quantum spin.
\end{thebibliography}

\end{document}